\documentclass[useAMS,usenatbib]{mn2e}

\usepackage[active]{srcltx}
\usepackage{graphicx}
\usepackage{txfonts}
\usepackage{natbib}
\usepackage{multirow}
\usepackage{longtable}
\usepackage[applemac]{inputenc}
\usepackage{rotating}

\usepackage{color}
\usepackage{soul}
\definecolor{jaune}{rgb}{1.0, 1.0, 0.0}

\newcommand*{\vcenteredhbox}[1]{\begingroup
\setbox0=\hbox{#1}\parbox{\wd0}{\box0}\endgroup}

\newcommand{\kms}{km\, s$^{-1}\,$}

\def\gtrsim{\mathrel{\hbox{\rlap{\hbox{\lower4pt\hbox{$\sim$}}}\hbox{$>$}}}}
\def\ltsim{\mathrel{\hbox{\rlap{\hbox{\lower4pt\hbox{$\sim$}}}\hbox{$<$}}}}

\title[The spectral variability and magnetic field characteristics of the Of?p star HD 148937]{The spectral variability and magnetic field characteristics\\of the Of?p star HD 148937\thanks{Based on observations obtained at the Canada-France-Hawaii Telescope (CFHT) which is operated by the National Research Council of Canada, the Institut National des Sciences de l'Univers of the Centre National de la Recherche Scientifique of France, and the University of Hawaii}}
\author[G.A. Wade et al.]{G.A. Wade$^{1}$, J. Grunhut$^{1,2}$, G. Gr\"afener$^3$, I.D. Howarth$^{4}$, F. Martins$^5$, V. Petit$^6$, 
\newauthor{J.S. Vink$^{3}$, S. Bagnulo$^3$, C.P. Folsom$^3$,Y. Naz\'e$^{7}$, N.R. Walborn$^{8}$, R.H.D. Townsend$^9$, } 
\newauthor{C.J. Evans$^{10}$ and the MiMeS Collaboration}\thanks{E-mail: wade-g@rmc.ca}\\
$^1$Dept. of Physics, Royal Military College of Canada, PO Box 17000, Stn Forces, Kingston, Ontario K7K 7B4, Canada \\
$^2$Dept. of Physics, Engineering Physics and Astronomy, Queen's University, 99 University Avenue, Kingston, Ontario K7L 3N6, Canada\\
$^3$Armagh Observatory, College Hill, Armagh, Northern Ireland, UK BT61 9DG\\
$^4$Dept. of Physics and Astronomy, UCL, Gower Place, London WC1E 6BT, United Kingdom \\
$^5$LUPM-UMR5299, CNRS \& Universit\'e Montpellier II, Place Eug\`ene Bataillon, F-34095, Montpellier, France\\
$^6$Department of Geology \& Astronomy, West Chester University, West Chester, Pennsylvania, USA 19383\\
$^{7}$FNRS-Institut d'Astrophysique et de G\'eophysique, Universit\'e de Li\`ege, Belgium\\
$^8$Space Telescope Science Institute, 3700 San Martin Drive, Baltimore, MD 21218, USA\\
$^9$Dept. of Astronomy, University of Wisconsin-Madison, 475 N. Charter Street, Madison WI 53706-1582, USA\\
$^{10}$UK ATC, Royal Observatory Edinburgh, Blackford Hill, Edinburgh, EH9 3HJ, UK}

\begin{document}

\date{Accepted . Received ; in original form }

\pagerange{\pageref{firstpage}--\pageref{lastpage}} \pubyear{2002}

\maketitle

\label{firstpage}

\begin{abstract}
We report magnetic and spectroscopic observations and modeling of the Of?p star HD 148937 within the context of the Magnetism in Massive Stars (MiMeS) Large Program at the Canada-France-Hawaii Telescope. Thirty-two high signal-to-noise ratio circularly polarised (Stokes $V$) spectra and 13 unpolarised (Stokes $I$) spectra of HD 148937 were acquired in 2009 and 2010. A definite detection of a Stokes $V$ Zeeman signature is obtained in the grand mean of all observations (in both Least-Squares Deconvolved (LSD) mean profiles and individual spectral lines). The longitudinal magnetic field inferred from the Stokes $V$ LSD profiles is consistently negative, in contrast to the essentially zero field strength measured from the diagnostic null profiles. A period search of new and archival equivalent width measurements confirms the previously-reported $7.03$~d variability period.  The variation of equivalent widths is not strictly periodic: we present evidence for evolution of the amount or distribution of circumstellar plasma. Interpreting the 7.03d period as the stellar rotational period within the context of the Oblique Rotator paradigm, we have phased the equivalent widths and longitudinal field measurements. The longitudinal field measurements show a weak sinusoidal variation of constant sign, with extrema out of phase with the H$\alpha$ variation by about 0.25 cycles. From our constraint on $v\sin i\leq 45$~km\,s$^{-1}$, we infer that the rotational axis inclination $i\leq 30\degr$. Modeling the longitudinal field phase variation directly, we obtain the magnetic obliquity $\beta=38^{+17}_{-28}\degr$ and dipole polar intensity $B_{\rm d}=1020^{-380}_{+310}$~G. Simple modeling of the H$\alpha$ equivalent width variation supports the derived geometry. The inferred magnetic configuration confirms the suggestion of Naz\'e et al (2010), who proposed that the weaker variability of HD 148937 as compared to other members of this class is a consequence of the stellar geometry. Based on the derived magnetic properties and published wind characteristics, we find a wind magnetic confinement parameter $\eta_*\simeq 20$ and rotation parameter $W=0.12$, supporting a picture in which the H$\alpha$ emission and other line variability have their origin in an oblique, rigidly rotating magnetospheric structure resulting from a magnetically channeled wind.\end{abstract}



%
%

\section{Introduction}

The enigmatic Of?p stars are identified by a number of peculiar and outstanding observational properties. The classification was first introduced by Walborn (1972) according to the presence of C~{\sc iii} $\lambda 4650$ emission with a strength comparable to the neighbouring N~{\sc iii} lines. Well-studied Of?p stars are now known to exhibit recurrent, and apparently periodic, spectral variations (in Balmer, He~{\sc i}, C~{\sc iii} and Si~{\sc iii} lines), narrow P Cygni or emission components in the Balmer lines and He~{\sc i} lines, and UV wind lines weaker than those of typical Of supergiants (see Naz\'e et al. 2010 and references therein). 

Only 5 Galactic Of?p stars are known (Walborn et al. 2010): HD 108, HD 148937, HD 191612, NGC 1624-2 and CPD$-28\degr 2561$. Three of these stars - HD 108, HD 148937 and HD 191612 - have been studied in detail. In recent years, HD 191612 and HD 108 have been carefully examined for the presence of magnetic fields (Donati et al. 2006; Martins et al. 2010; Wade et al 2011), and both have been clearly detected. In retrospect, the similarity of some of the observational properties of the Of?p stars to the O7 V magnetic oblique rotator $\theta^1$~Ori C appears to have been a strong indicator that they are also oblique rotators with strong, stable, organised magnetic fields. As the brightest known Of?p star, and in light of the report of the marginal detection of a longitudinal magnetic field by Hubrig et al. (2008, 2011), the time is ripe for a more intensive investigation of HD 148937.

HD~148937 has been extensively observed; its observational properties are discussed in the literature by Naz\'e et al. (2008a, 2010). Similar to other well-studied Of?p stars, HD 148937 is revealed to be a high mass (50-60~$M_\odot$) main sequence O-type star with an effective temperature of about 40~kK. The physical parameters of this star (as first derived by Naz\'e et al. 2008a and revised in this paper) are reported in Table 1. The HR diagram positions of the 3 well-studied Of?p stars are illustrated in Fig. 1. While it is clear that the Of?p stars are high-mass O-type stars, their HR diagram positions are relatively uncertain due to their poorly constrained distances. This translates into important absolute and relative uncertainties on their masses and ages, as is reflected in the relative large error bars in Fig. 1. 

Like HD 108 and HD 191612, HD~148937 is a spectroscopic variable star. However, it is distinguished from those stars by its lower-amplitude variability, particularly in the primary diagnostic H$\alpha$ line. Whereas HD 108 and HD 191612 exhibit H$\alpha$ equivalent width (EW) variability of more than a factor of 2, the H$\alpha$ line of HD 148937 changes in EW by only about 20\% (e.g. Naz\'e et al. 2008a). Despite this much weaker variability, Naz\'e et al. (2008a) were able to tentatively identify a period of $7.031\pm 0.003$~d in EW measurements of HD 148937 spanning over 3 years. This period was subsequently confirmed using higher quality data by Naz\'e et al. (2010). Photometry of HD 148937 reveals no significant broadband flux variation (Balona 1992, van Genderen et al. 1989, 2001, Naz\'e et al. 2008a).

Naz\'e et al. (2008a) also found that this star exhibits an enhanced nitrogen abundance (by about a factor of four relative to the sun). At X-ray wavelengths, HD 148937 resembles HD 108 and HD 191612 in having a thermal spectrum dominated by a relatively cool component ($kT=0.2$~keV), broad lines ($>1700$~km\,s$^{-1}$; although see detection of narrower features in high-$Z$ lines by Naz\'e et al. 2011), and an order-of-magnitude overluminosity compared to normal O stars ($\log [L_{\rm X}^{\rm unabs}/L_{\rm BOL}] \sim -6$).

In this paper we perform a first detailed investigation of the combined magnetic and variability properties of HD 148937 using an extensive high-resolution spectropolarimetric and spectroscopic dataset. In Sect. 2 we discuss the data acquired and the methods of analysis used. In Sect. 3 we re-examine the physical properties of the star, as well as its projected rotational velocity. In Sect. 4 we examine the spectral variability, confirming the periodic variability of the H$\alpha$ and other emission and absorption lines, and the period derived by Naz\'e et al. (2008a, 2010). In Sect. 5 we analyse in detail the magnetic data acquired at the Canada-France-Hawaii Telescope (CFHT). In Sect. 6 we employ the CFHT magnetic data, in combination with the stellar physical and rotational parameters, to constrain the magnetic field strength and geometry. Finally, in Sect. 7 we explore the implications of our study, particularly regarding the variability and other properties of HD 148937, the confinement and structure of its stellar wind, and of the properties of the general class of Of?p stars.

\begin{figure}
\centering
\includegraphics[width=8cm,angle=0]{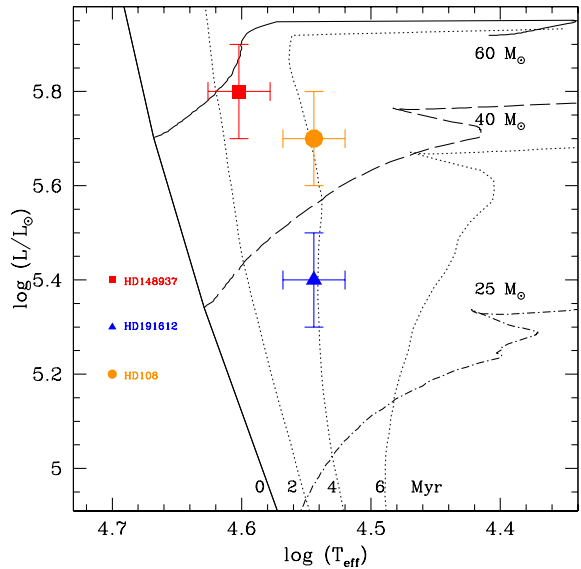}
\caption{HR diagram showing the positions of the 3 Of?p stars in which magnetic fields have been detected: HD 108, HD 191612 and HD 148937. According to its position on this diagram, HD 148937 is the youngest, and most massive and luminous of the well-studied Of?p stars (although it should be noted that the detailed positions are subject to significant uncertainties in the distances to these stars, as reflected in the error bars). Evolution tracks and isochrones include rotational mixing, and are from Meynet \& Maeder (2005). Adapted from Martins et al. (2010).}
\label{SNRs}
\end{figure}





\begin{table}
\centering
\caption{Summary of stellar physical, wind and magnetic properties of HD~148937, derived by Naz\'e et al. (2008a) and in this paper. The wind magnetic confinement parameter $\eta_*$, the rotation parameter $W$ and the characteristic spin down time $\tau_{\rm spin}$ are defined and described in Sect. 7.}
\begin{tabular}{l|ll}
\hline
Spectral type & Of?p    &          \\
$T_{\rm eff}$ (K) & 41 000 $\pm$ 2000 & Naz\'e et al. (2008a)\\
log $g$ (cgs) & 4.0 $\pm$ 0.1  & Naz\'e et al. (2008a)   \\
R$_{\star}$ (R$_\odot$) & 15.0 $\pm 2.5$ & Naz\'e et al. (2008a)\\
$v\sin i$ (km\,s$^{-1}$) & $\ltsim 45$ & This paper  \\
$\log (L_\star/L_\odot)$ & 5.8 $\pm$ 0.1  & This paper\\
$M_{\star}$ ($M_{\odot}$) & $\sim 60$ & This paper\\
\hline
$\log \dot{M}/\sqrt{f}$ (M$_{\odot}$\,yr$^{-1}$) & $-6$  & Naz\'e et al. (2008a)\\
$v_{\infty}$ (km\,s$^{-1}$) & 2600  & Naz\'e et al. (2008a)\\
Fill factor $f$ & 0.01 & Naz\'e et al. (2008a)\\
\hline
$B_{\rm d}$ (G) & $1020^{-380}_{+310}$ & This paper\\
$\beta$ ($\degr$) & $38^{+17}_{-28}$ & This paper\\
\hline
$\eta_*$ & 20 & This paper\\
$W$ & $1.2\times 10^{-1}$ & This paper\\
$\tau_{\rm spin}$ & 1.6 Myr & This paper\\
\hline
$E(B-V)$ & 0.67 & This paper\\
$\log(L_X/L_{Bol})$ & -6.1 & Naz\'e et al. (2008a)\\
\hline\hline
\end{tabular}
\label{params}
\end{table}

\section{Observations}

\subsection{ESPaDOnS spectropolarimetry}

Spectropolarimetric observations of HD 148937 were obtained using the ESPaDOnS spectropolarimeter at the CFHT in 2009 and 2010 within the context of the Magnetism in Massive Stars (MiMeS) Large Program and Project. Altogether, 32 Stokes $V$ sequences were obtained - two in May 2009, two in September 2009, 14 during 7 consecutive nights in June 2010, and 14 during 7 consecutive nights in July 2010.  In particular, the final observing runs were planned so as to sample two complete rotations of the star, hypothesising that the 7.03 d period identified by Naz\'e et al. was in fact the stellar rotational period.

Each polarimetric sequence consisted of four individual subexposures taken in different polarimeter configurations. From each set of four subexposures we derive a mean Stokes $V$ spectrum following the procedure of Donati et al. (1997), ensuring in particular that all spurious signatures are removed at first order. Diagnostic null polarization spectra (labeled $N$) are calculated by combining the four subexposures in such a way that polarization cancels out, allowing us to check that no spurious signals are present in the data (see Donati et al. 1997 for more details on how $N$ is defined). All frames were processed using the Upena pipeline feeding Libre ESpRIT (Donati et al. 1997), a fully automatic reduction package installed at CFHT for optimal extraction of ESPaDOnS spectra. The peak signal-to-noise ratios per 1.8 km\,s$^{-1}$ velocity bin in the reduced spectra range from 60 to nearly 1000, with a median of 690, depending on the exposure time and on weather conditions. Due to the low declination of HD 148937, the star was generally observed at high airmass ($2.8-3.5$) from CFHT.

The log of CFHT observations is presented in Table 2.

\begin{table*}
\caption{Log of CFHT MiMeS observations of HD 148937, including results of the magnetic analysis described in Sect. 3. S/N is peak signal-to-noise ratio per 1.8~\kms\ pixel in the reduced spectrum (occurring around H$\alpha$). Phase is computed using the ephemeris expressed by Eq. (2). $z=B_{\rm z}/\sigma$ is the detection significance of the longitudinal magnetic field.}
\begin{center}
\begin{tabular}{cccccrrrrccccccc}\hline\hline
            &                   &                   &             &             & \multicolumn{2}{c}{Stokes $V$} & \multicolumn{2}{c}{Null $N$}\\
CFHT & HJD          & Exposure & Peak  & Phase & $\langle B_{\rm z}\rangle \pm \sigma_{\rm B}$ & $z_V$ & $\langle B_{\rm z}\rangle \pm \sigma_{\rm B}$ & $z_N$\\
Odometer & -2450000           & time (s)    &        S/N           &                             &    (G)      &                    &(G) &                                                          \\
\hline
1075687 &2454955.04670   &2720 &927 & 0.100  &  $  -81  \pm 101   $  & -0.8     &      $   96 \pm 101 $   &  1.0 \\  
1076968 &2454959.97316   &2720 &529 & 0.800  &  $ -312  \pm 187   $  & -1.7     &      $ -278 \pm 187 $   & -1.5 \\  
1115108 &2455078.72811   &2400 &532 & 0.687  &  $ -242  \pm 187   $  & -1.3     &      $  243 \pm 187 $   &  1.3 \\  
1115112 &2455078.75798   &2400 &516 & 0.692  &  $  -72  \pm 205   $  & -0.4     &      $  180 \pm 204 $   &  0.9 \\  
1206040 &2455366.87889   &2400 &780 & 0.663  &  $ -440  \pm 118   $  & -3.7     &      $  262 \pm 116 $   &  2.3 \\  
1206044 &2455366.90878   &2400 &764 & 0.667  &  $ -188  \pm 126   $  & -1.5     &      $  242 \pm 125 $   &  1.9 \\  
1206203 &2455367.88596   &2400 &764 & 0.806  &  $ -247  \pm 122   $  & -2.0     &      $   66 \pm 120 $   &  0.6 \\  
1206207 &2455367.91586   &2400 &721 & 0.810  &  $ -308  \pm 132   $  & -2.3     &      $   13 \pm 132 $   &  0.1 \\  
1206344 &2455368.85844   &2400 &700 & 0.944  &  $  -21  \pm 139   $  & -0.2     &      $    5 \pm 137 $   &  0.0 \\  
1206348 &2455368.88837   &2400 &725 & 0.949  &  $ -541  \pm 135   $  & -4.0     &      $ -224 \pm 135 $   & -1.7 \\  
1206558 &2455369.88524   &2400 &506 & 0.090  &  $ -214  \pm 196   $  & -1.1     &      $  218 \pm 198 $   &  1.1 \\  
1206562 &2455369.91519   &2400 &505 & 0.095  &  $ -645  \pm 200   $  & -3.2     &      $ -190 \pm 199 $   & -1.0 \\  
1206741 &2455370.84461   &2400 &389 & 0.227  &  $ -112  \pm 245   $  & -0.5     &      $  347 \pm 246 $   &  1.4 \\  
1206745 &2455370.87453   &2400 &421 & 0.231  &  $ -229  \pm 228   $  & -1.0     &      $  235 \pm 227 $   &  1.0 \\  
1206893 &2455371.84232   &2400 &760 & 0.369  &  $ -235  \pm 119   $  & -2.0     &      $  -97 \pm 117 $   & -0.8 \\  
1207032 &2455371.87222   &2400 &765 & 0.373  &  $ -293  \pm 128   $  & -2.3     &      $  -22 \pm 127 $   & -0.2 \\  
1207036 &2455372.84803   &2400 &967 & 0.512  &  $  -90  \pm 124   $  & -0.7     &      $   53 \pm 123 $   &  0.4 \\  
1206897 &2455372.87793   &2400 &782 & 0.516  &  $ -201  \pm 123   $  & -1.6     &      $   63 \pm 123 $   &  0.5 \\  
1217152 &2455400.77099   &2400 &721 & 0.482  &  $ -213  \pm 129   $  & -1.7     &      $   78 \pm 129 $   &  0.6 \\  
1217156 &2455400.80176   &2400 &648 & 0.487  &  $ -280  \pm 145   $  & -1.9     &      $  258 \pm 145 $   &  1.8 \\  
1217445 &2455401.77634   &2400 &58  & 0.625  &  $ 2246  \pm 2710   $  &  0.8     &      $ 178 \pm 2739$   &  0.1 \\  
1217450 &2455401.81689   &2400 &193 & 0.631  &  $ -658  \pm 551   $  & -1.2     &      $  282 \pm 550 $   &  0.5 \\  
1217692 &2455402.77201   &2400 &251 & 0.767  &  $ -465  \pm 409   $  & -1.1     &      $ -163 \pm 409 $   & -0.4 \\  
1217696 &2455402.80787   &2400 &317 & 0.772  &  $ -160  \pm 319   $  & -0.5     &      $  576 \pm 321 $   &  1.8 \\  
1217846 &2455403.77822   &2400 &587 & 0.910  &  $   24  \pm 168   $  &  0.1     &      $  223 \pm 165 $   &  1.4 \\  
1217850 &2455403.80871   &2400 &582 & 0.914  &  $  -58  \pm 173   $  & -0.3     &      $ -174 \pm 175 $   & -1.0 \\  
1218069 &2455404.78463   &2400 &677 & 0.053  &  $ -105  \pm 138   $  & -0.8     &      $ -328 \pm 138 $   & -2.4 \\  
1218073 &2455404.81527   &2400 &648 & 0.057  &  $ -169  \pm 139   $  & -1.2     &      $ -101 \pm 138 $   & -0.7 \\  
1218262 &2455405.76066   &2400 &728 & 0.192  &  $ -156  \pm 128   $  & -1.2     &      $   44 \pm 127 $   &  0.3 \\  
1218266 &2455405.79129   &2400 &797 & 0.196  &  $  -61  \pm 113   $  & -0.5     &      $  140 \pm 114 $   &  1.2 \\  
1218487 &2455406.75036   &2400 &744 & 0.333  &  $  -76  \pm 123   $  & -0.6     &      $  -92 \pm 123 $   & -0.8 \\  
1218491 &2455406.78074   &2400 &786 & 0.337  &  $   20  \pm 115   $  &  0.2     &      $ -143 \pm 113 $   & -1.3 \\\hline\hline\end{tabular}
\end{center}
\end{table*}

\subsection{FEROS spectroscopy}

In addition to the 32 circular polarisation spectra, 13 high-resolution unpolarised spectra were acquired on 10 nights
in 2009 March using the FEROS spectrograph on the MPG/ESO 2.2-m
telescope at La Silla.   These spectra cover a useful wavelength range of
$\sim$360--920 nm at a resolving power of $R \simeq 48,000$, with a S/N of
$\sim$125 at H$\alpha$.   The \'echelle data were extracted and merged
using the pipeline FEROS Data Reduction System (see Kaufer et
al. 1999 for details).   The log is given in Table 3.

\subsection{Coralie spectroscopy}

Finally, we have re-analysed the 20 \'echelle spectra of HD 148937 obtained in 2008 with the Coralie spectrograph on the 1.2m Euler Swiss telescope at La Silla (Chile), described by Naz\'e et al. (2010).

\section{Physical and rotational properties}

We have re-determined the main stellar properties of HD~148937 using the atmosphere code CMFGEN (Hillier \& Miller 1998). The results are basically unchanged compared to the analysis of Naz\'e et al. (2008a). The only minor difference is a slightly higher luminosity ($\log L/L_{\odot} = 5.8$ vs. 5.75). We consider this to be a better determination, based on the fit of the UV-optical-IR SED rather than just on the $V$-band magnitude. We also derive the colour excess $E(B-V)=0.67$ from this fitting process, assuming a classical Galactic extension law (Seaton 1979, Howarth 1983). Fits illustrating our results are presented in Fig. 2.


\begin{figure*}
\centering
\vcenteredhbox{\includegraphics[width=11cm]{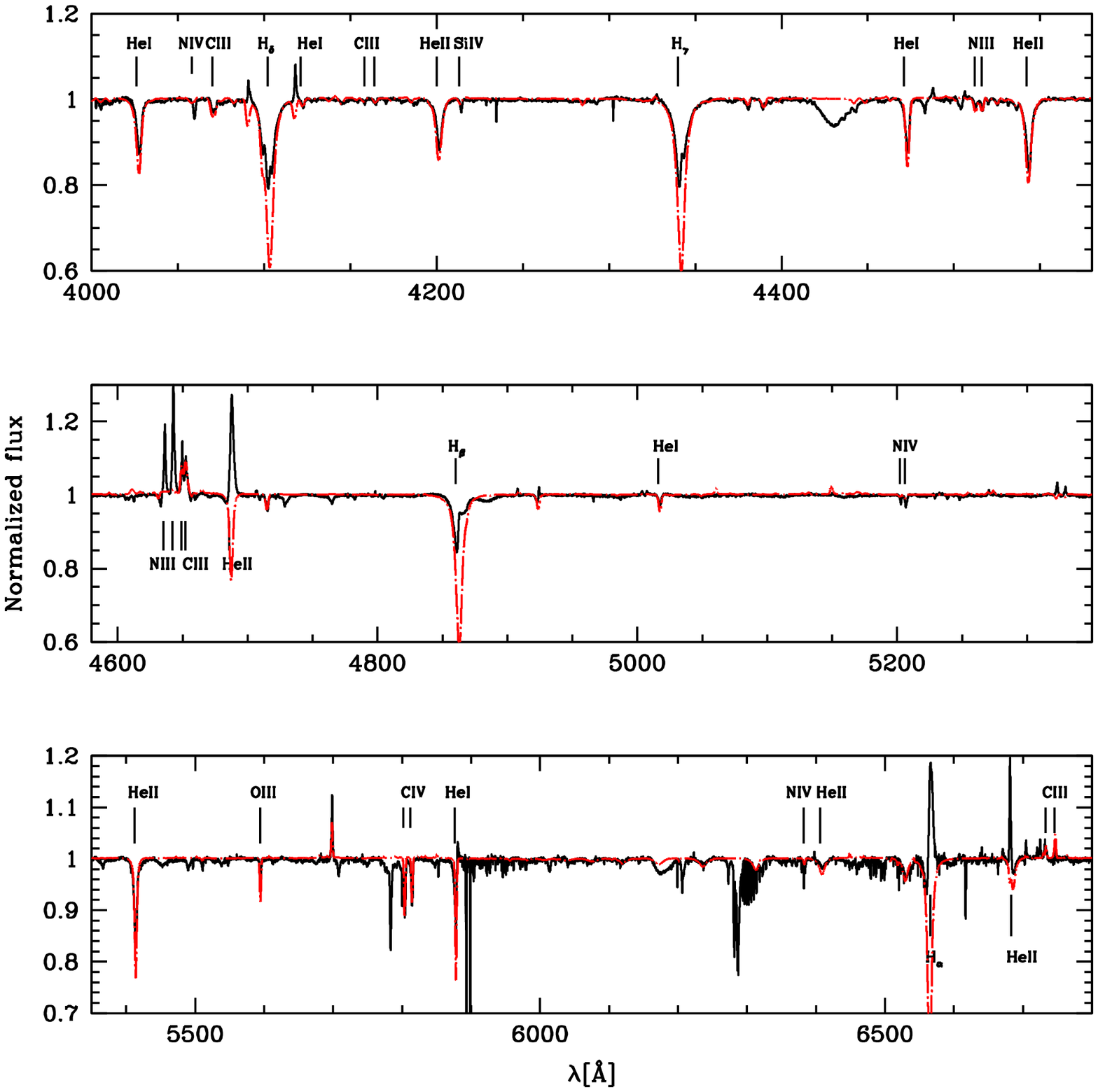}}\vcenteredhbox{\includegraphics[width=7cm]{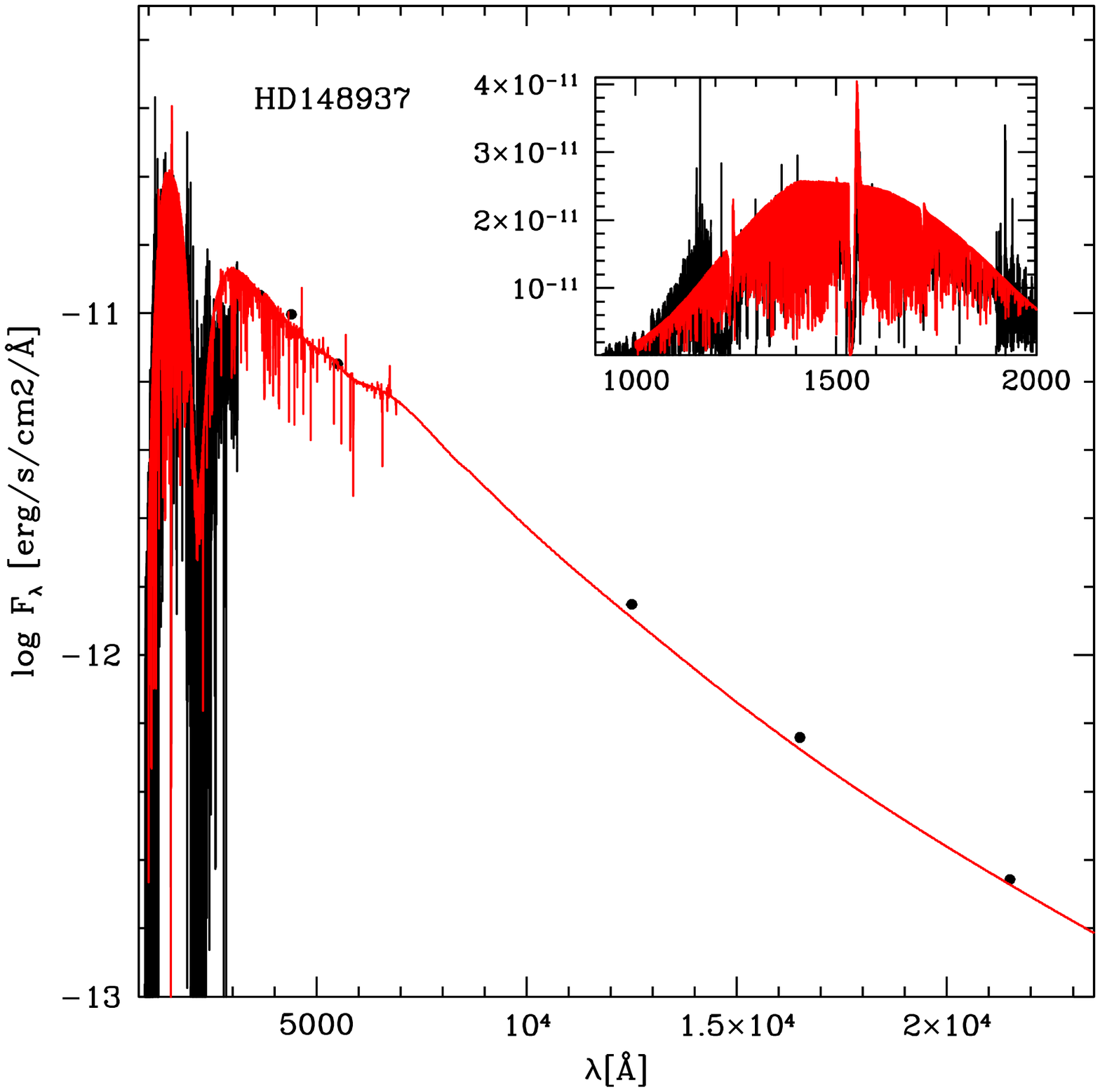}}
\caption{{\em Left panel: }\ Fit to the optical spectrum \#1206203. The black solid line is the observed spectrum, the dashed red line the best fit model ($T_{\rm eff}=40000$~K, $\log g=4.0$, $\log L/L_{\odot}=5.8$, $\dot{M}=10^{-7}~M_{\odot}$/yr, $v_{\infty}=2600$~km\,s$^{-1}$). The Balmer lines are not reproduced by the TLUSTY models. (see Sect. 3).  {\em Right panel:}\ Fit of the UV-optical-IR SED. The red solid line is the best fit model ($\log L/L_{\odot}=5.8$ and $E(B-V)=0.67$). The black solid line is the FUSE-IUE spectrum. The black points correspond to $UBVJHK$ photometry.}
\label{SNRs}
\end{figure*}

\begin{table}
\caption{Log of Feros spectroscopic measurements of HD 148937. S/N is the inferred signal-to-noise ratio per pixel based on scatter on continuum pixels near H$\alpha$. Phase is computed using the ephemeris expressed by Eq. (2). }
\begin{center}
\begin{tabular}{crrrrrrrrrrrrrrr}\hline\hline
HJD & Phase  &   $t_{\rm exp}$ & S/N & \\
\hline
2454904.8818  & 0.966 &60 & 138 \\
2454905.9261  & 0.115 &60 & 144 \\
2454907.9267  & 0.399 &60 & 119 \\
2454908.9197  & 0.541 &60 & 124 \\
2454909.8465  & 0.672 &60 & 102 \\
2454910.9096  & 0.824 &60 & 115 \\
2454910.9109  & 0.824 &60 & 125 \\
2454911.9068  & 0.965 &60 & 101 \\
2454912.8990  & 0.106 &60 & 139 \\
2454913.8248  & 0.238 &60 & 118 \\
2454913.8262  & 0.238 &60 & 123 \\
2454914.9080  & 0.392 &60 & 130 \\
2454914.9093  & 0.392 &60 & 131 \\\hline\hline
\end{tabular}
\end{center}
\end{table}

\begin{figure*}
\centering
\includegraphics[width=5cm,angle=-90]{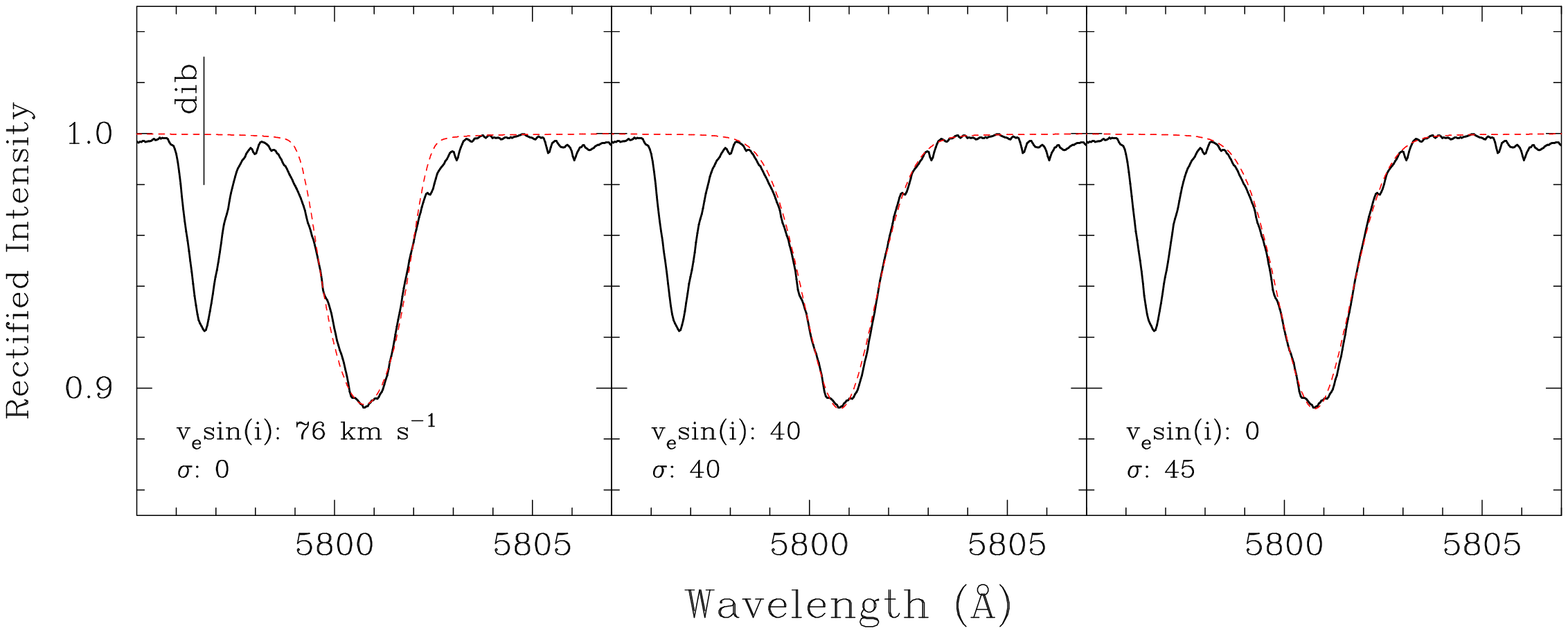}
\caption{Fits to the C~{\sc iv} $\lambda 5801$ line profile for illustrative combinations of rotation and macroturbulence.  The left-hand panel shows that a rotationally broadened model matching the line width fails to match the line shape; an additional broadening mechanism is required.  The remaining panels show that the projected rotational velocity is poorly constrained, because of this significant turbulent broadening.}
\label{SNRs}
\end{figure*}

Based on our spectral analysis and comparison with the line profiles of HD 108 and HD 191612, we conclude that macroturbulence is likely the dominant line broadening mechanism. Tests performed using He~{\sc i} $\lambda$4713 and C~{\sc iv} $\lambda\lambda$5801/5812 reveal that convolution of our synthetic spectra with a Gaussian profile (mimicking isotropic macroturbulence) gives better results compared to pure rotational broadening. In practice, a macroturbulence of about 50~km\,s$^{-1}$ gives satisfactory fits. However, many profiles - in particular those of He~{\sc i} - have a strong (wind-induced?) asymmetry, and those that have only small asymmetries have broadening that is dominated by 'turbulence'. The projected rotational velocity is an important parameter needed to constrain the stellar and magnetic geometry. To place a stronger constraint on the $v\sin i$ of HD 148937, we focused mainly on the C~{\sc iv}~$\lambda 5801$ line as it is reasonably strong, and reasonably symmetric.   

We used a 40~kK, $\log=4.0$ TLUSTY (Lanz \& Hubeny 2003) model for the
intrinsic line profiles, slightly scaled to match the observed line
strength.  The model profiles were rotationally broadened with a
simple convolution and a linear limb-darkening coefficient of 0.4 (numerical integration of specific intensities
gives results that are indistinguishable); isotropic Gaussian
turbulence; and a Gaussian instrumental response, before remapping to
the sampling rate of the observations, and adding matching noise.
Comparison of models \& observations was performed for a
range of $v\sin i$ and turbulence, in both wavelength and Fourier space.

We find that the broadening is roughly described by the constraint that

\begin{equation}
\sqrt{\zeta^2 + (v\sin i)^2} \simeq\ {\rm 100\ km\,s}^{-1},
\end{equation}

\noindent where $\zeta$ is the FWHM (2.354$\sigma$) of the Gaussian turbulence. Interestingly, this is essentially identical to
the result for HD 191612 (Howarth et al. 2007). This is remarkable, given the large difference in their inferred rotational periods. As
suggested by Wade et al. (2011), this supports the view that rotational broadening is a negligible contributor to the observed line widths.

The upper limit to $v\sin i$ from C~{\sc iv}~$\lambda 5801$ is 60 km\,s$^{-1}$, and $v\sin i=0$ provides an acceptable fit (with $\zeta=106$~km\,s$^{-1}$). This is illustrated in Fig. 3.  
We also explored a variety of weaker lines, but even with such high S/N spectra we are limited by
S/N issues.  Nevertheless, in the end we concluded that the N~{\sc iv}~$\lambda 5226$ line gives a slightly
better upper limit on $v\sin i$ of about 45 km\,s$^{-1}$.

These numbers are compatible with those reported by Naz\'e et al. (2008a).  The properties of HD 148937 as derived in this section are summarised in Table 1.

\begin{figure}
\centering
\includegraphics[width=7cm,angle=-90]{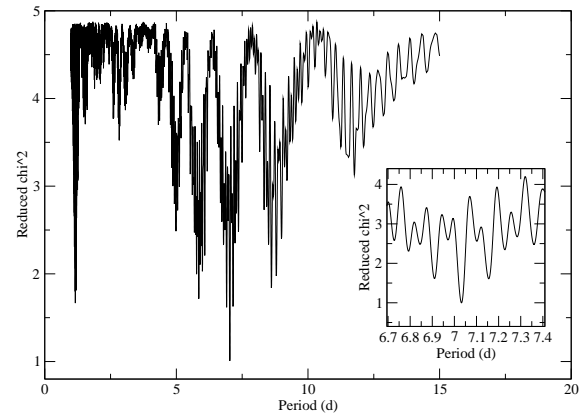}\caption{Combined periodogram for all data sets yielding periods near 7.03 d. The adopted period is $7.032\pm 0.003$~d.}
\label{EWs}
\end{figure}

\begin{figure*}
\centering

\includegraphics[width=5cm]{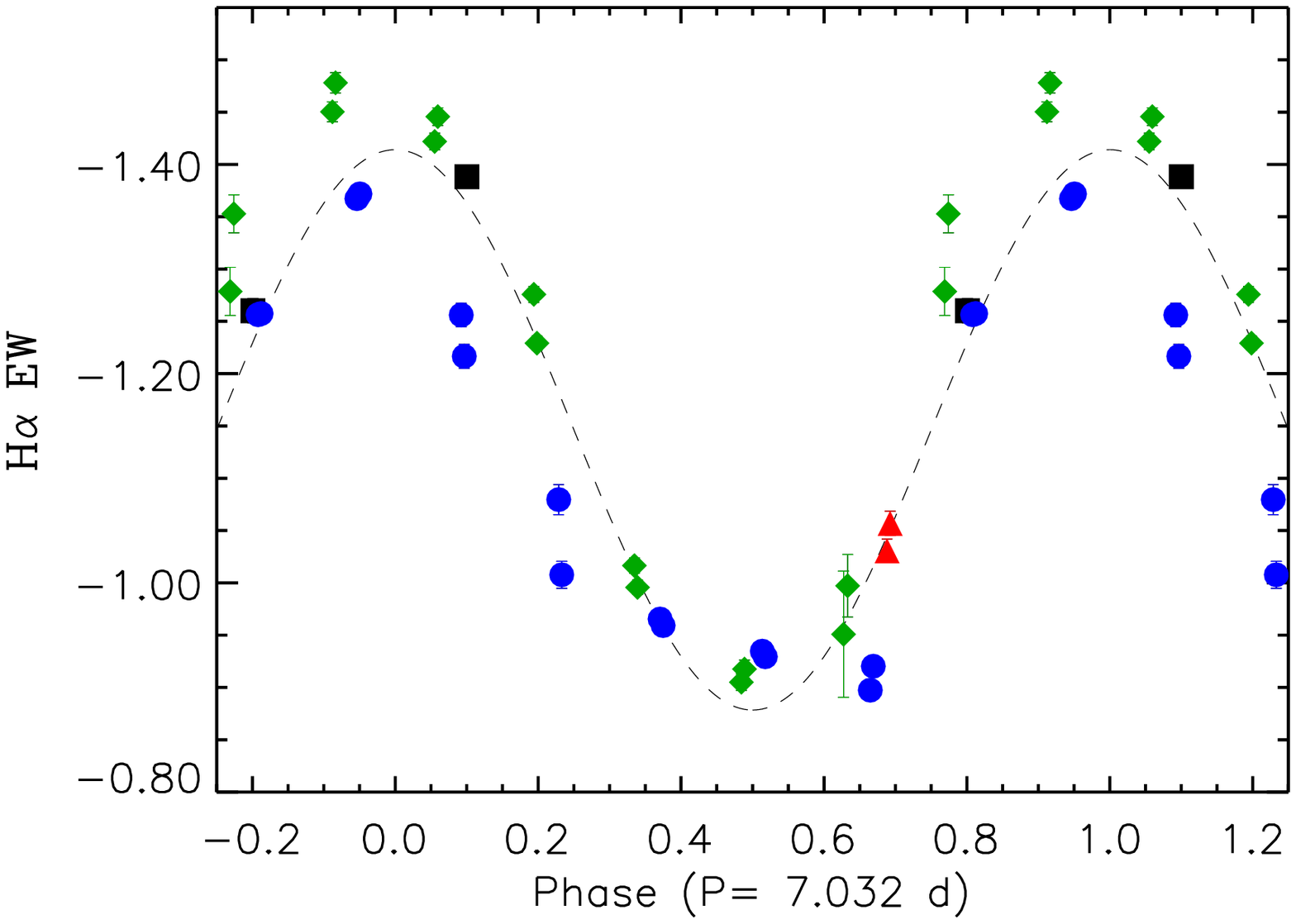}\includegraphics[width=5cm]{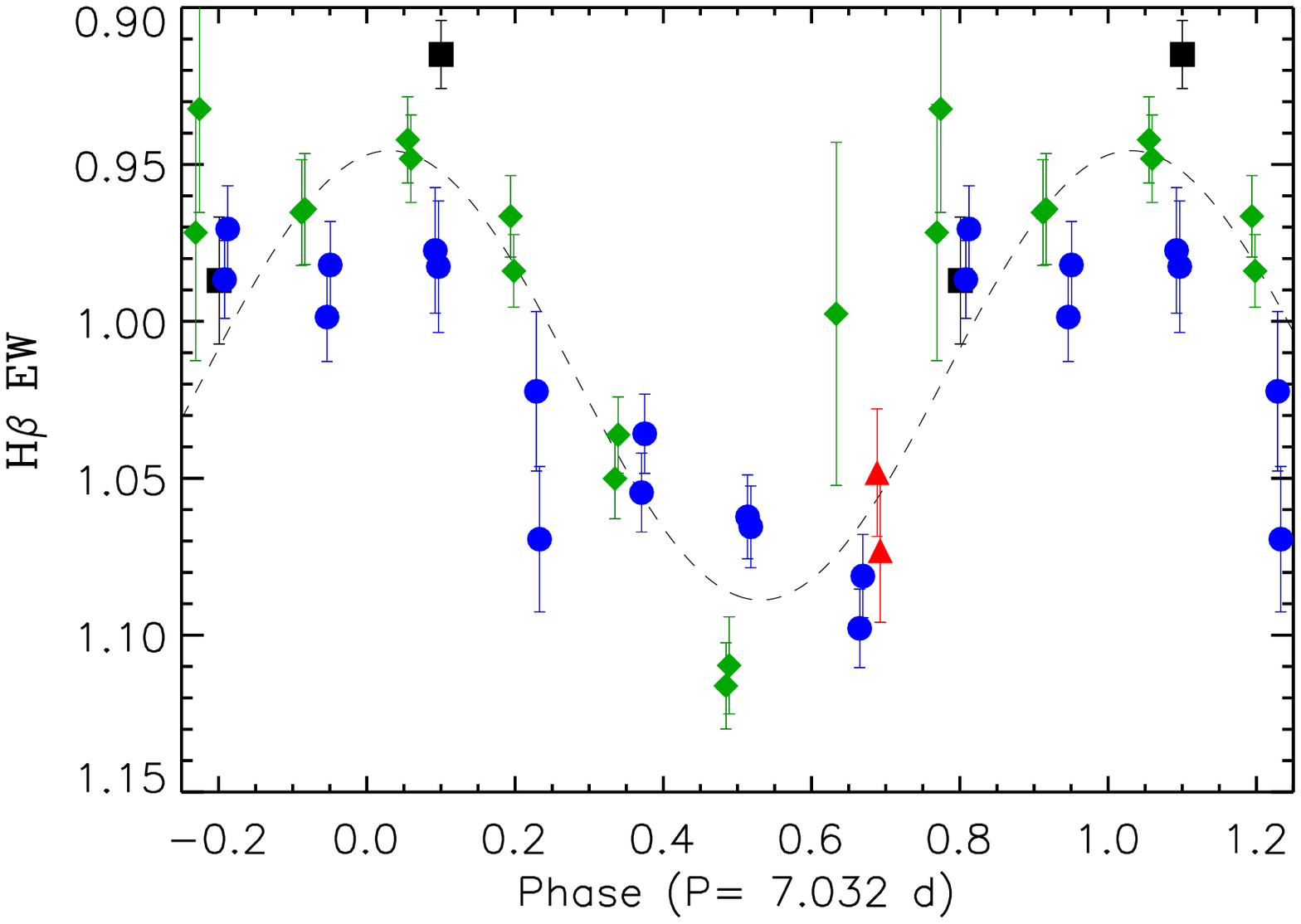}\includegraphics[width=5cm]{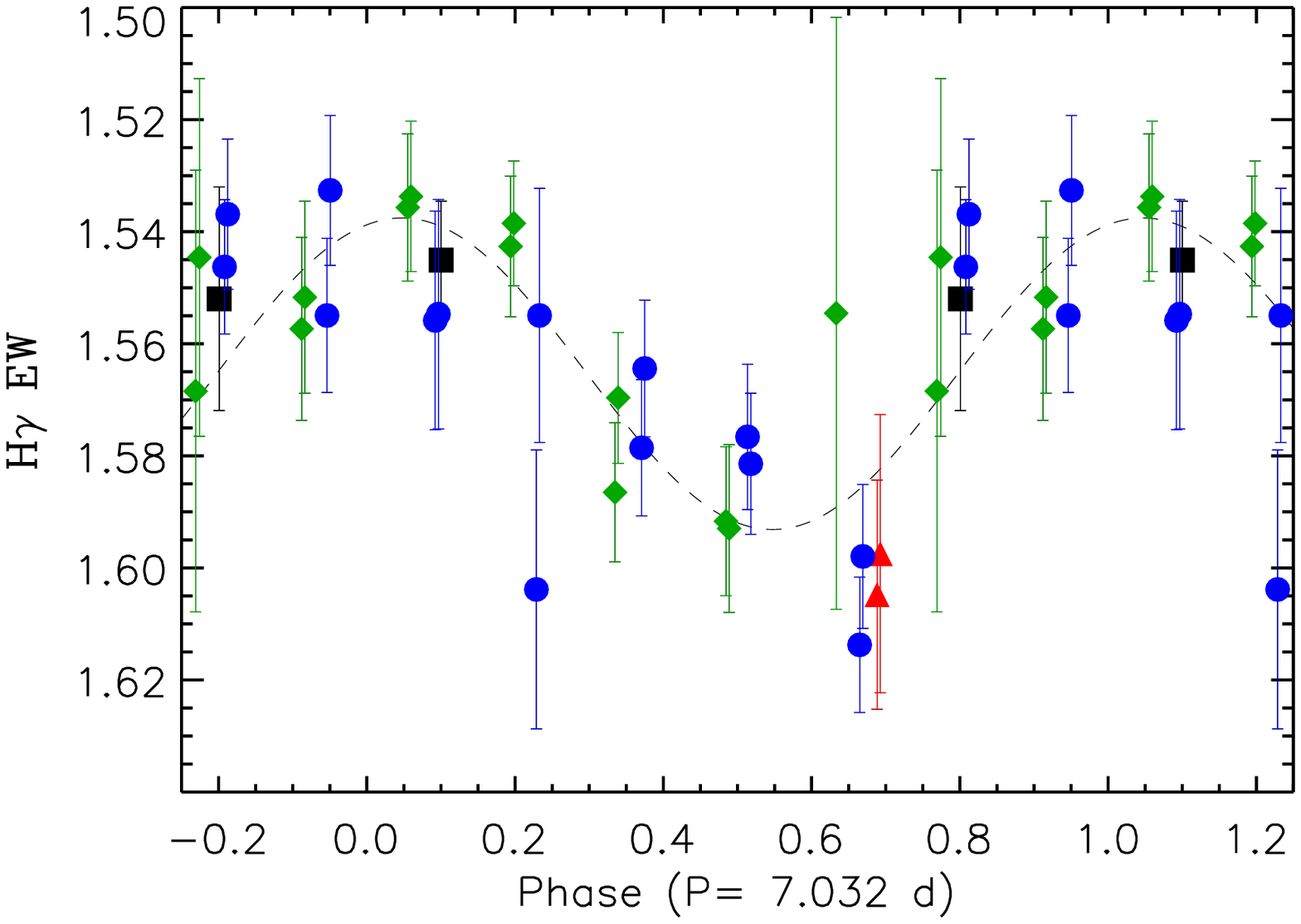}

\includegraphics[width=5cm]{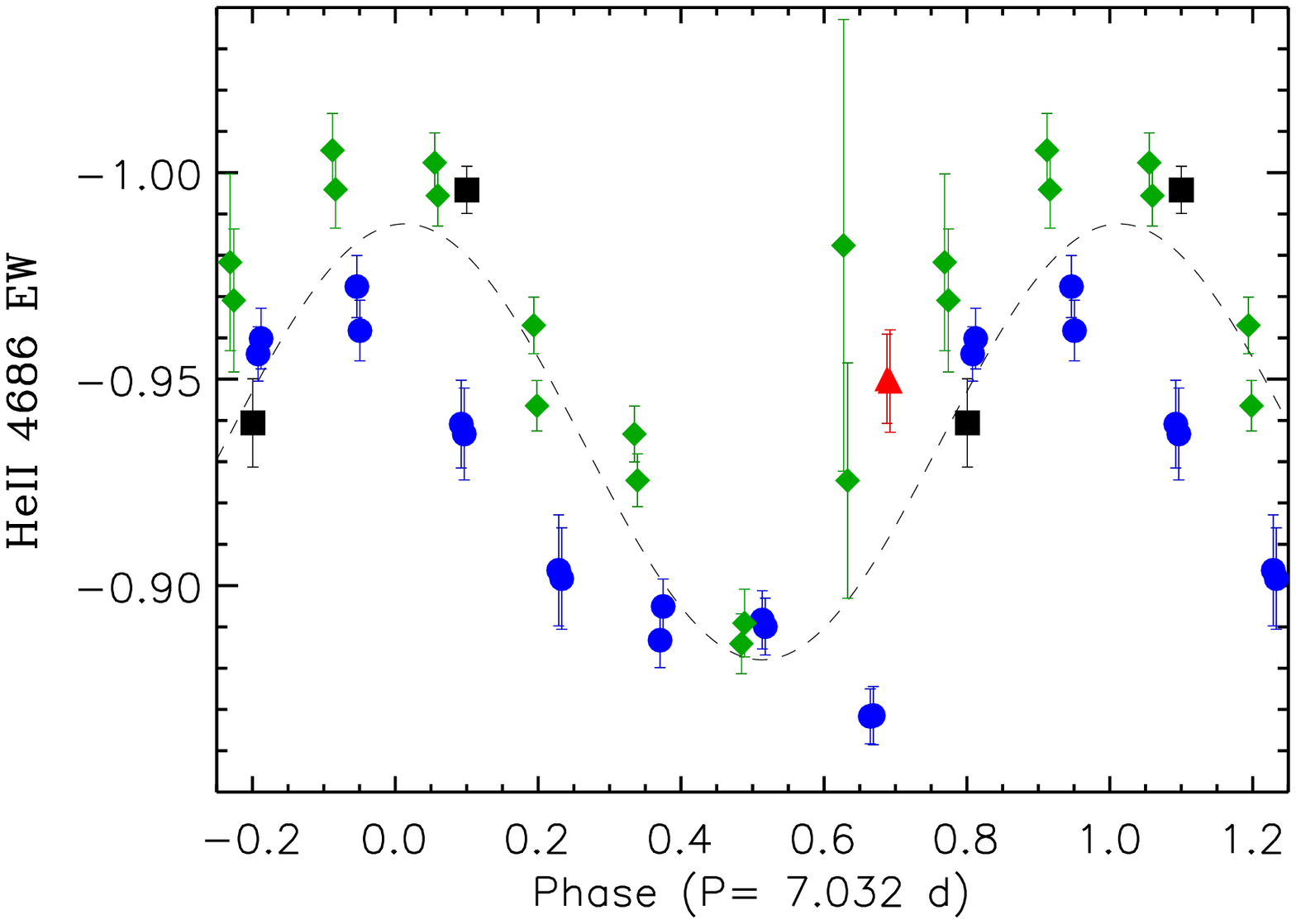}\includegraphics[width=5cm]{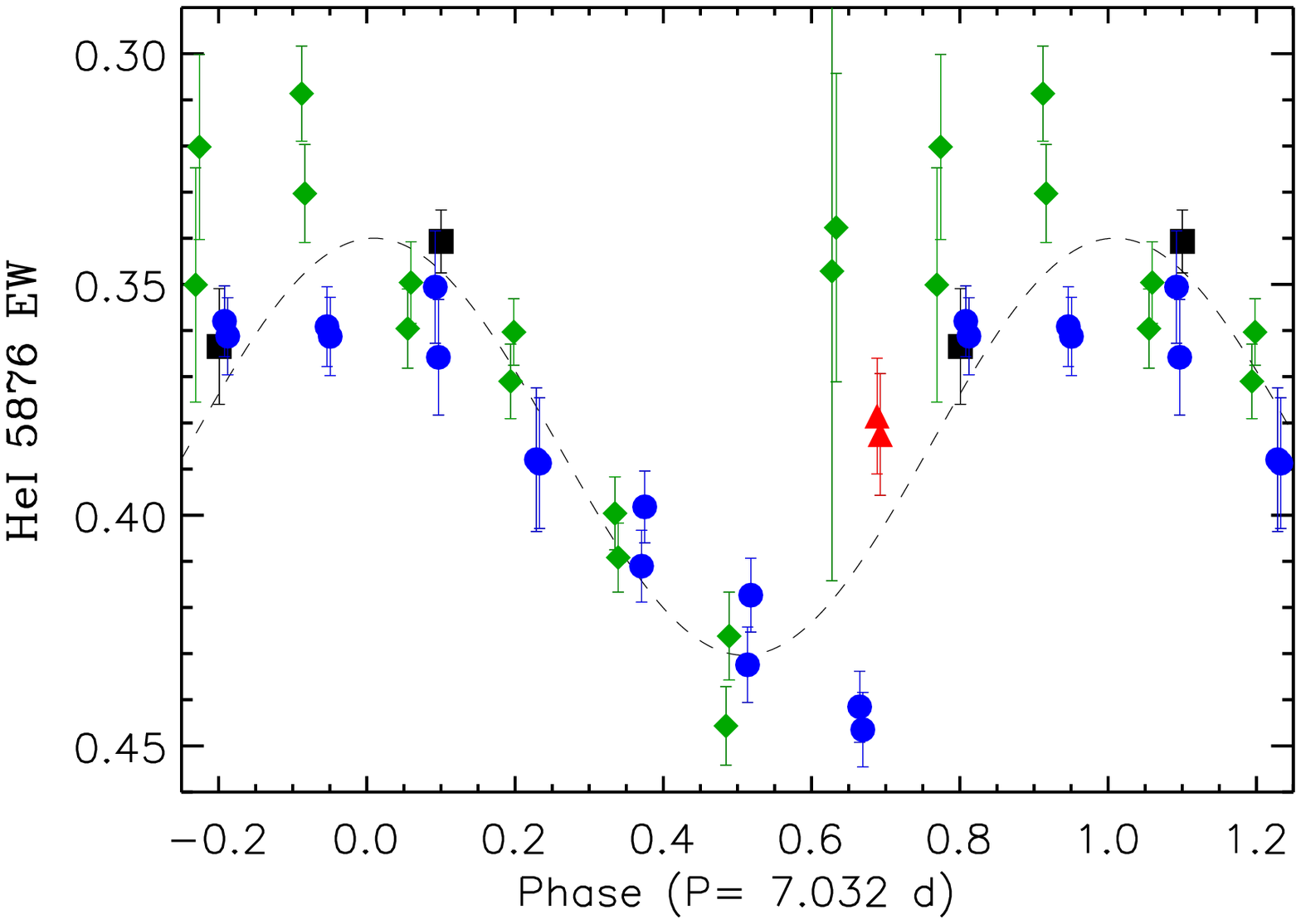}\includegraphics[width=5cm]{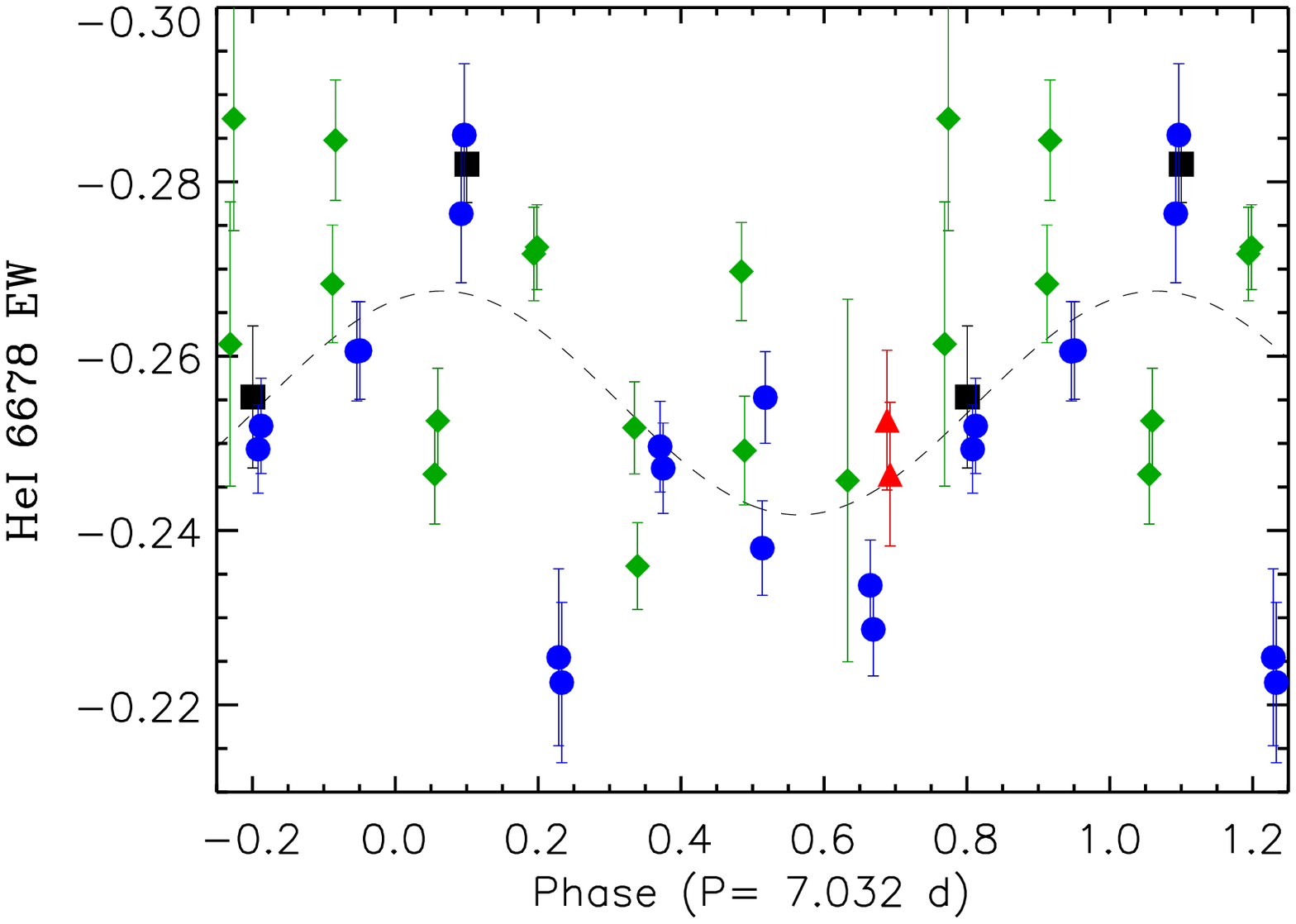}

\includegraphics[width=5cm]{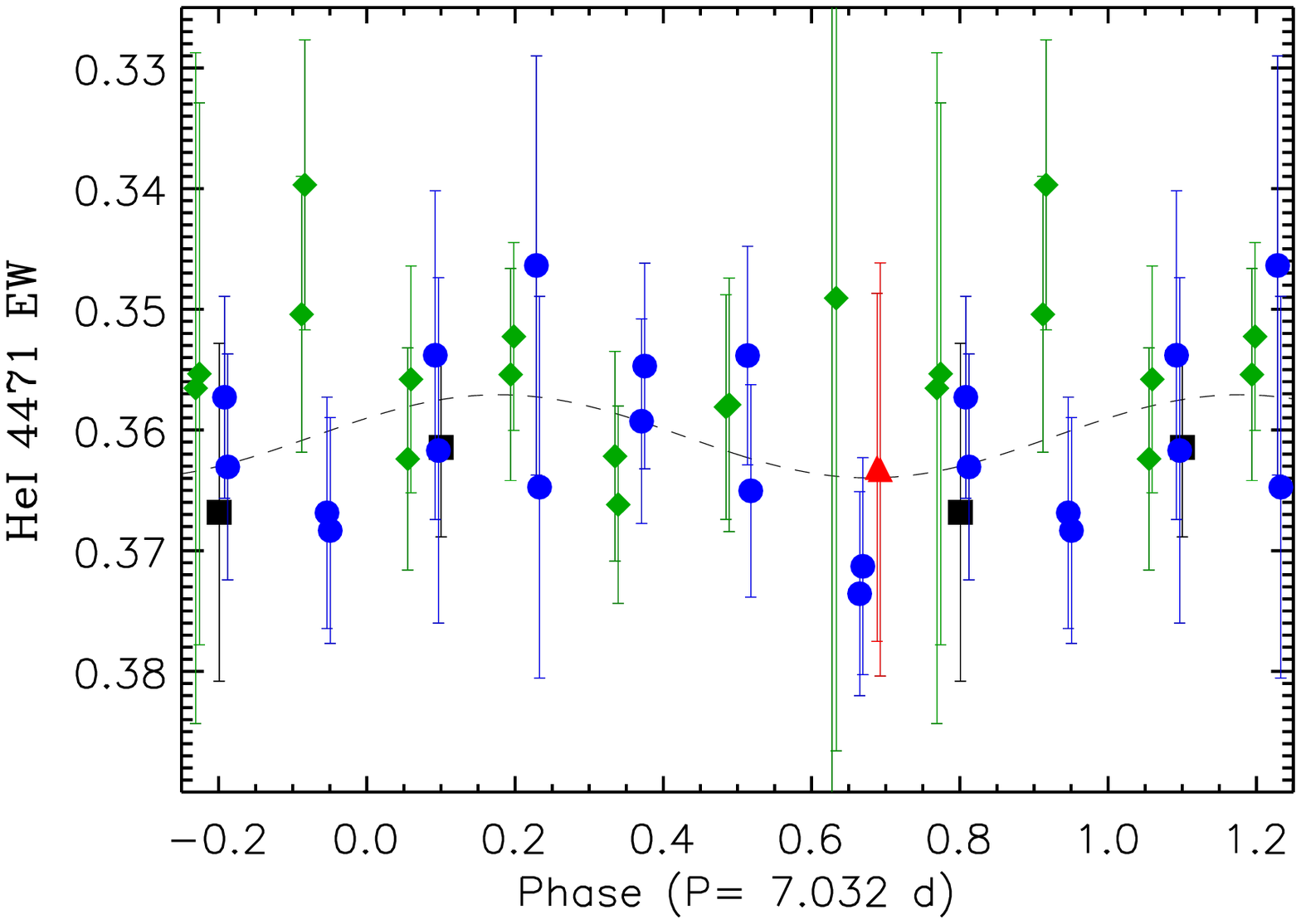}\includegraphics[width=5cm]{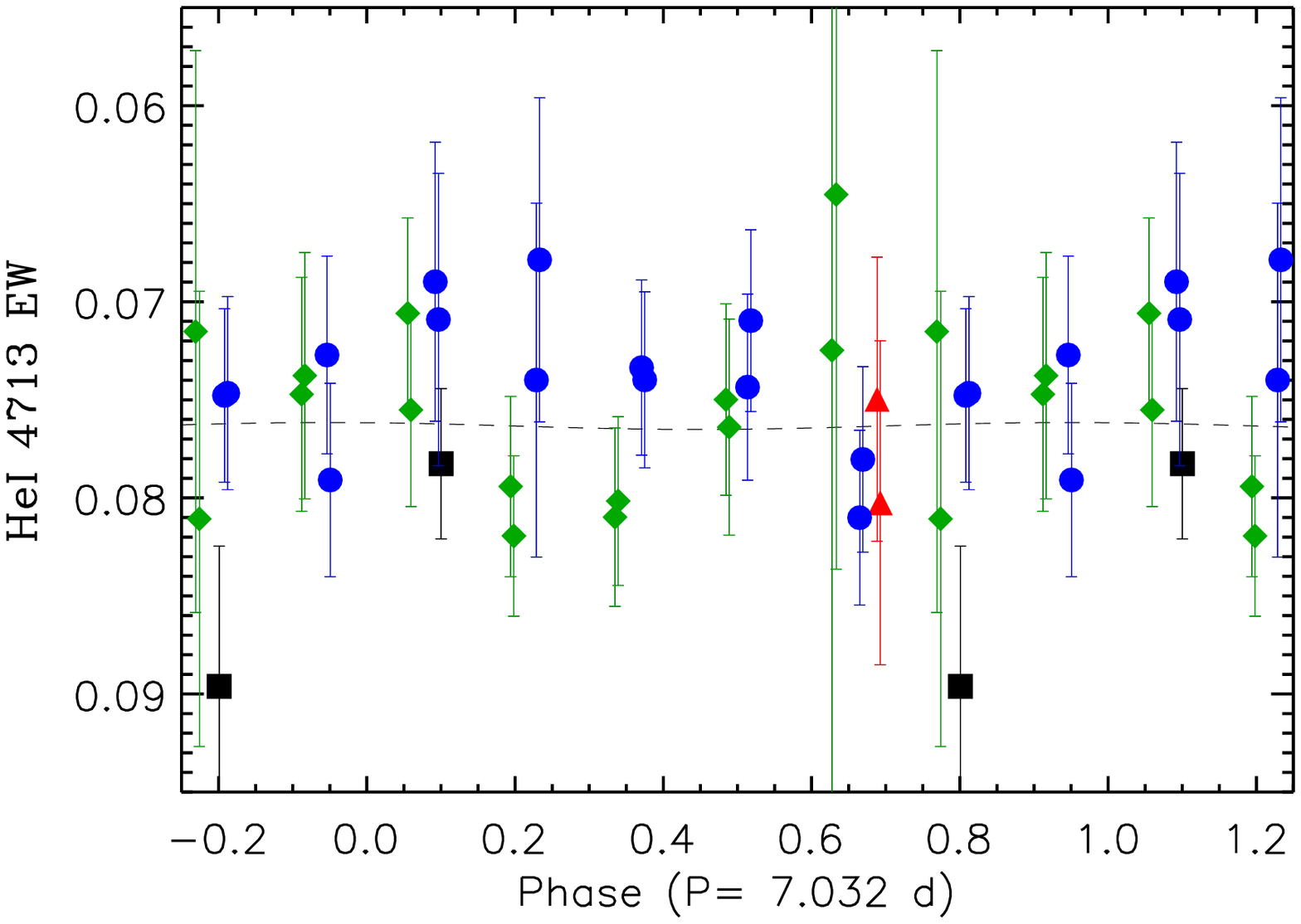}\includegraphics[width=5cm]{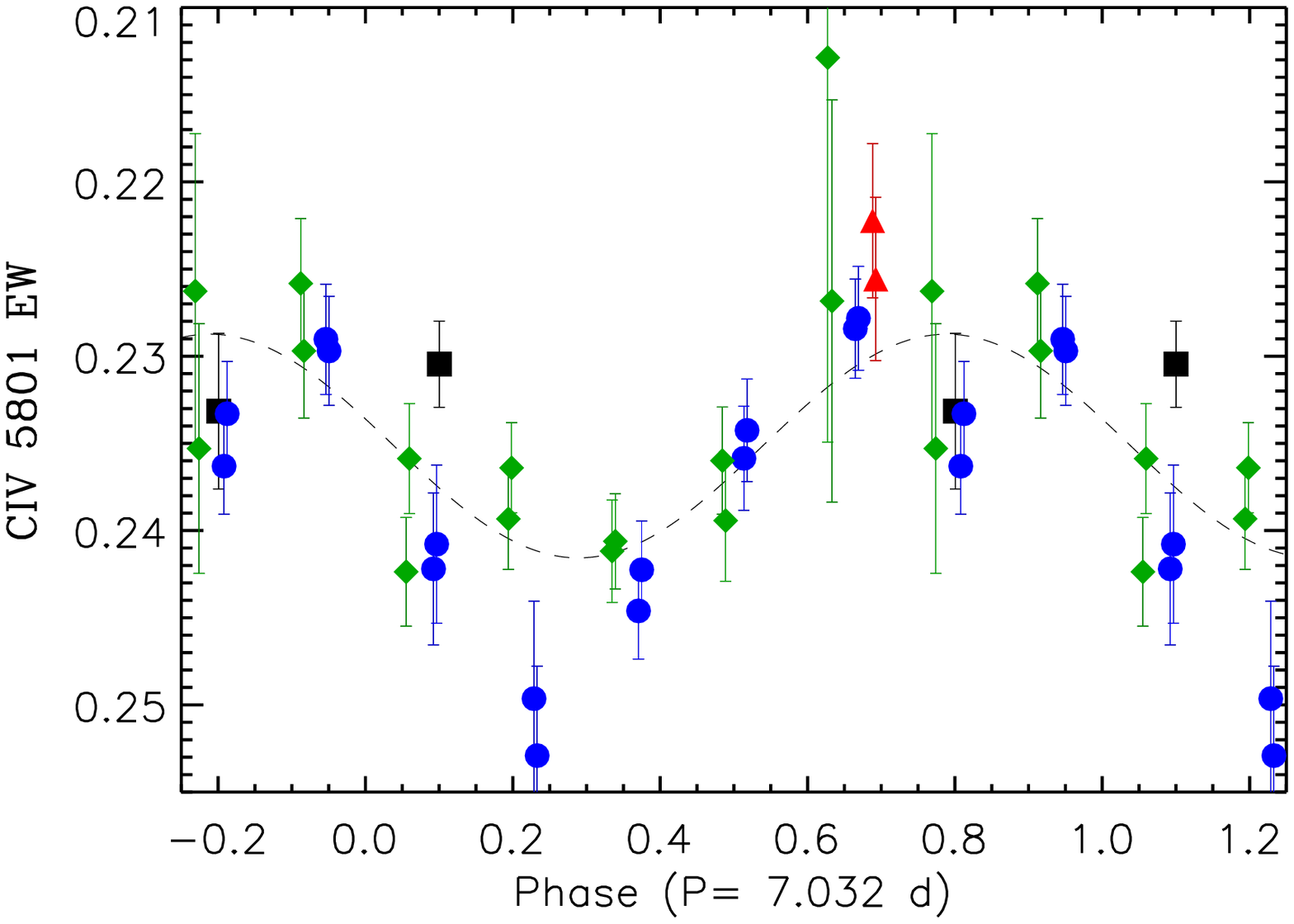}

\includegraphics[width=5cm]{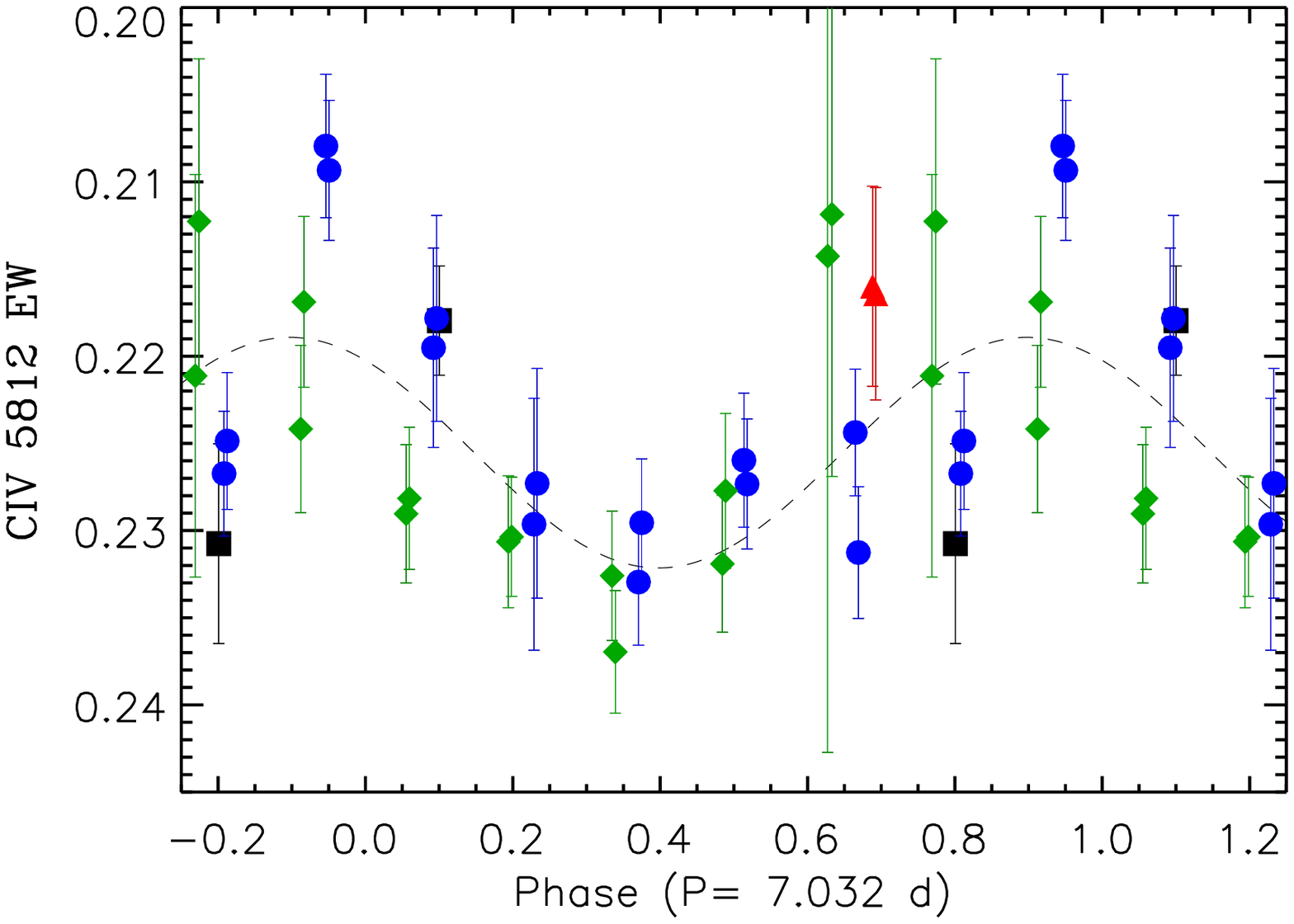}\includegraphics[width=5cm]{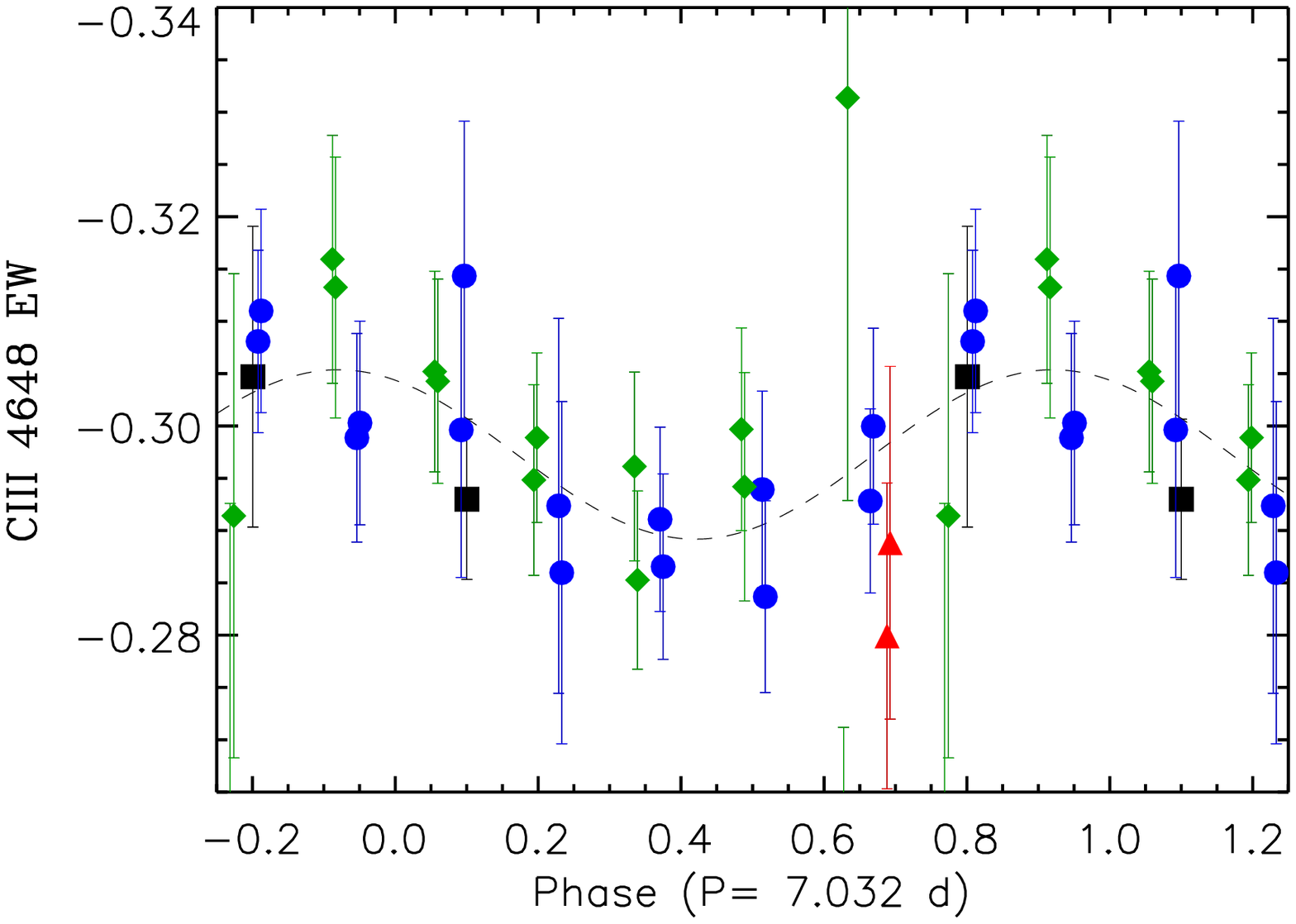}\includegraphics[width=5cm]{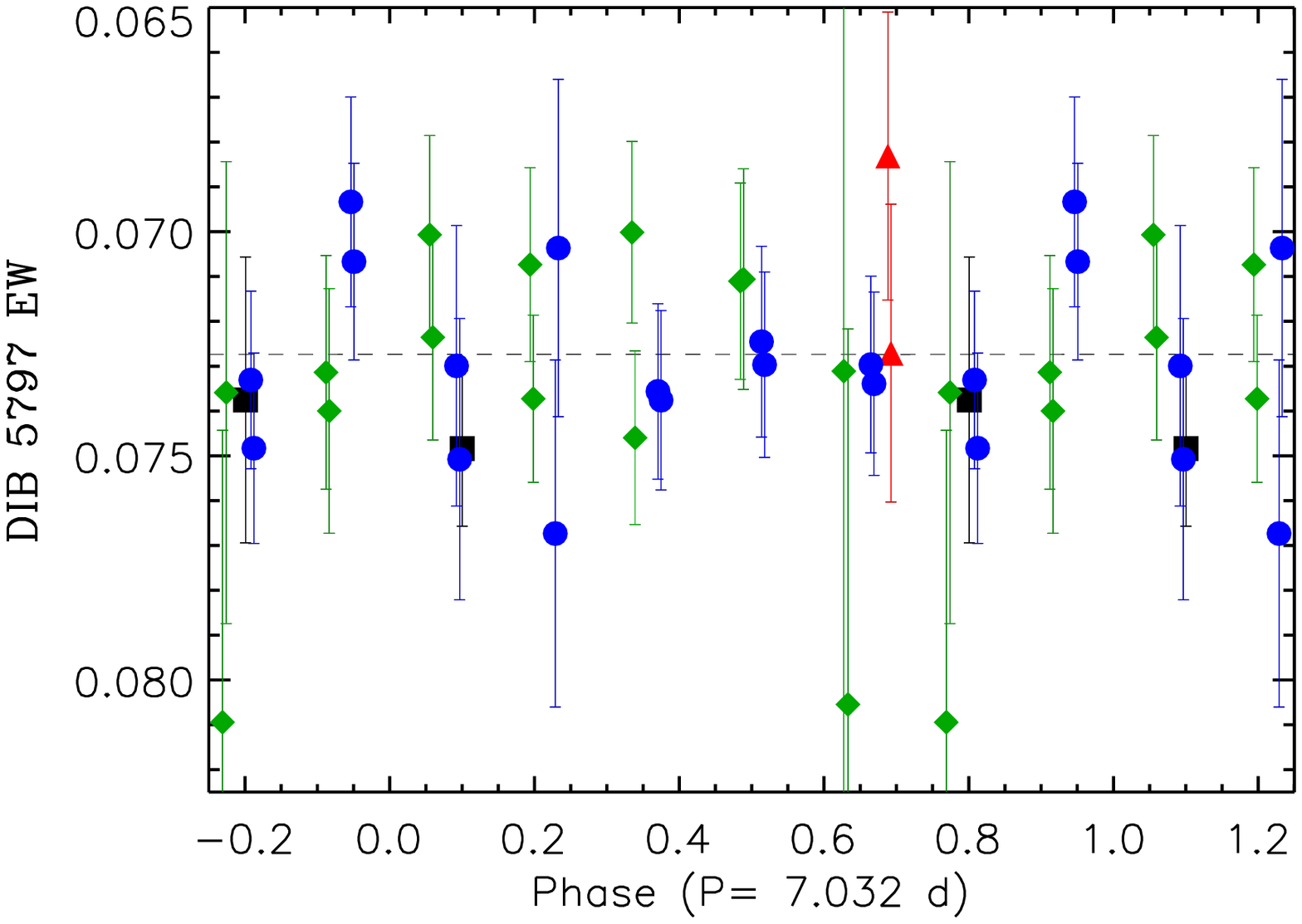}

\caption{Phased EW variations of lines in the ESPaDOnS spectra of HD 148937: Balmer lines H$\alpha$, H$\beta$ and H$\gamma$; He~{\sc ii} $\lambda$4686, He~{\sc i} $\lambda$5876,  He~{\sc i} $\lambda$6678; He~{\sc i} $\lambda$4471, He~{\sc i} $\lambda$4713; C~{\sc iv} $\lambda 5801$, C~{\sc iv} $\lambda 5812$; C~{\sc iii} $\lambda 4650$ lines; and the DIB located at 5797~\AA.  Symbols/colours distinguish different cycles. Dashed curves represent least-squares sinusoidal fits to the combined ESPaDOnS data.}
\label{EWs}
\end{figure*}

\section{Spectral variability}

As described by e.g. Naz\'e et al. (2008a), the complex spectrum of HD 148937 exhibits many similarities to HD 108 and HD 191612, as well as some important differences. Like HD 108 and HD 191612, the C~{\sc iii}~$\lambda 4650$ lines are in emission, with strength comparable to the nearby N lines. However, according to Naz\'e et al., unlike those stars, the C~{\sc iii} lines are not observed to vary. In fact, due to the substantially higher S/N of the new CFHT data, we detect weak variability in a large number of spectral lines, including the C~{\sc iii} lines.

{Following Naz\'e et al. (2008a, 2010) we characterise line variability using the equivalent width (EW). Before measuring the EW, each spectral line was locally re-normalized, and a telluric correction algorithm was applied to regions redward of 5790~\AA. The EWs were then obtained by numerically integrating over the line profile. The 1$\sigma$ uncertainties were calculated by propagating the individual pixel uncertainties in quadrature. Neither the Coralie nor the FEROS spectra had pre-computed uncertainties; therefore, a single uncertainty value was estimated from the RMS scatter in the continuum regions around the line profile and assigned to each pixel. 


The line showing the strongest variability is H$\alpha$, with a variability amplitude of 20-30\%. The variations of other photospherically-important lines showing distinct emission contributions - H$\gamma$, He~{\sc ii} $\lambda$4686, H$\beta$, He~{\sc i} $\lambda$5876 - are also statistically significant. He~{\sc i}~$\lambda 6678$ is an exception - this line is in full emission, but varies only weakly (the variation is detected only in the higher-S/N ESPaDOnS measurements, and then only marginally). These results are consistent with the report of Naz\'e et al. (2010).

Other lines of He~{\sc i} (e.g. $\lambda 4471, \lambda 4713$), while asymmetric, generally show no distinct emission. These lines show no significant variability. The C~{\sc iv} $\lambda 5801, \lambda 5812$ lines are slightly variable. This appears to be the first detection of EW variability of the optical C~{\sc iv} lines in an Of?p star.

We performed period searches using a Lomb-Scargle algorithm. As a first step, the procedure was applied to each of the timeseries of EW measurements of the strongly variable emission lines. We then used the method described by Folsom et al. (2008), treating each pixel within the line profile as an independent timeseries of measurements. Each line profile was interpolated onto the wavelength grid corresponding to one of the ESPaDOnS spectra. Periodograms were constructed for individual pixels in the profile. These periodograms were then averaged together, weighted by the amplitude of the variability in their respective pixels.  The resulting weighted mean periodogram for the profile was then examined for a best fit period. 

All periodograms of the emission line EWs show significant power near 7d. These periods, which are reproduced in the pixel-by-pixel periodograms with somewhat larger uncertainties, are all consistent with the best-fit H$\alpha$ period derived by Naz\'e et al. (2008a). To derive a single period characterising the variability of HD 148937, the periodograms resulting from the analysis of the EWs and line profile variations of the individual lines were all averaged. The final resulting periodogram, shown in Fig. 4, exhibits a single acceptable period at $7.032\pm 0.003$~d. This period is consistent with that reported by Naz\'e et al. (2010). Based on these results, for the remainder of the current study we adopt the following ephemeris for HD 148937:

 \begin{equation}
{\rm JD}_{H\alpha^{\rm max}}= 2454588.67(30)+7.032(3)\cdot E,
 \end{equation}
 
 \noindent where we have adopted a zero point corresponding to maximum H$\alpha$ emission as derived from a sinusoidal fit to the entire set of phased measurements (i.e. ESPaDOnS, FEROS and Coralie).

The phased EW measurements are illustrated in Fig. 5 (for illustrative clarity we show only the higher-quality ESPaDOnS spectra). The variations of all variable emission lines are in approximate phase with the H$\alpha$ line. Variability with a similar phasing is observed in the C~{\sc iii} 4650~\AA\ lines (which we note are predicted to be blended with an O~{\sc ii} line). Interestingly, the EW variations of the C~{\sc iv}~$\lambda 5801$ line appears to be shifted by approximately 0.25 cycles relative to the variations of the emission lines, although this shift is only marginally significant. We also include EWs measured from the DIB at 5797~\AA\ to illustrate the lack of variability of a reference line that is not formed in the environment of the star.


We find that in many cases the cycle-to-cycle scatter of the EWs is larger than the uncertainties attributable to noise, normalisation uncertainties and telluric line contamination. This appears to be a consequence of evolution of the shapes of the EW variations of some lines. This is especially evident in the H$\alpha$ and He~{\sc ii}~$\lambda 4686$ lines which exhibit clear changes in amplitude and shape of the EW variation between the two well-sampled cycles (illustrated as green diamonds and blue circles in Fig. 5). The cycle-to-cycle differences in the EW curves of these two lines track each other almost perfectly, and correspond to clear systematic differences in the profile shapes (Fig. 6) that are expressed in essentially all lines to some degree. As these lines are located throughout the spectrum, these differences cannot be a consequence of systematic errors in local normalisation. Nor, given the location of some lines (e.g. $\lambda 4686$) well into the blue part of the spectrum, can it be attributed to telluric lines. We therefore conclude that this variability is intrinsic to HD 148937. 

A potential explanation of these differences would be an error in the ephemeris expressed by Eq. (2). To check if this could be a plausible cause, we varied the period to see if we could improve the agreement between the EW variations from different epochs. We still find that the 7.032 d period yields the most consistent EW variations amongst different epochs and that changing the period within the uncertainty had a negligibly small effect on the relative phasing of the different epochs. We also compared the profiles from different epochs corresponding to similar EW values but with small relative phase offsets to see if the observed profiles are a better match than those with different EWs but essentially identical phases (e.g. as in Fig. 6). These comparisons show that we find larger discrepancies between the profiles with similar EW with small relative phase offsets than those shown in Fig. 6, confirming that adjusting the ephemeris is not a solution to this phenomenon.}

Based on these results, we conclude that the differences in the line profiles are best explained by intrinsic changes in the amount or distribution of emitting material with time, i.e. evolution of the magnetosphere of HD 148937. It is unclear if this evolution is secular or stochastic. This will be discussed further in Sect. 7.

\begin{figure*}
\centering
\includegraphics[width=5.5cm]{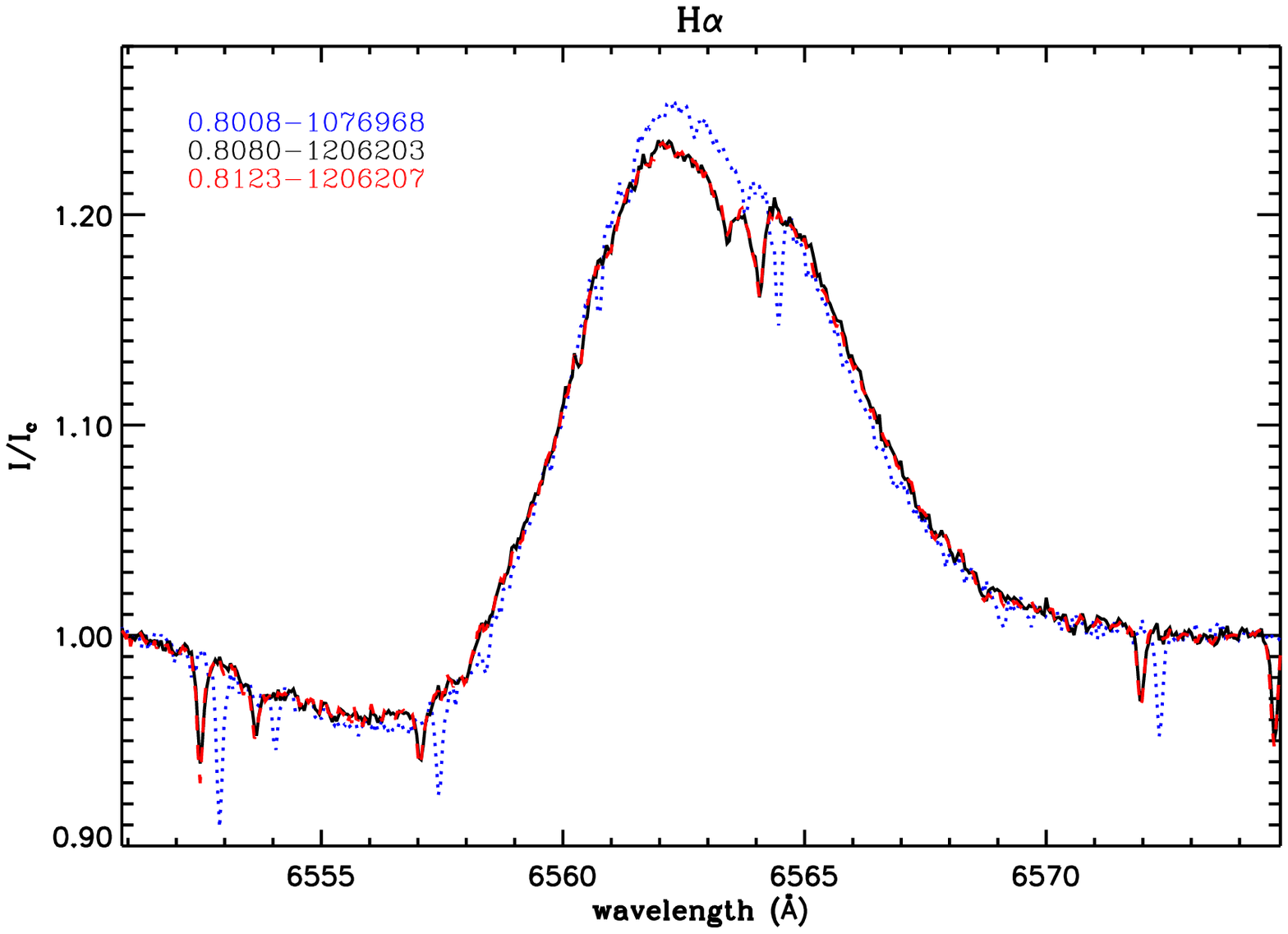}
\includegraphics[width=5.5cm]{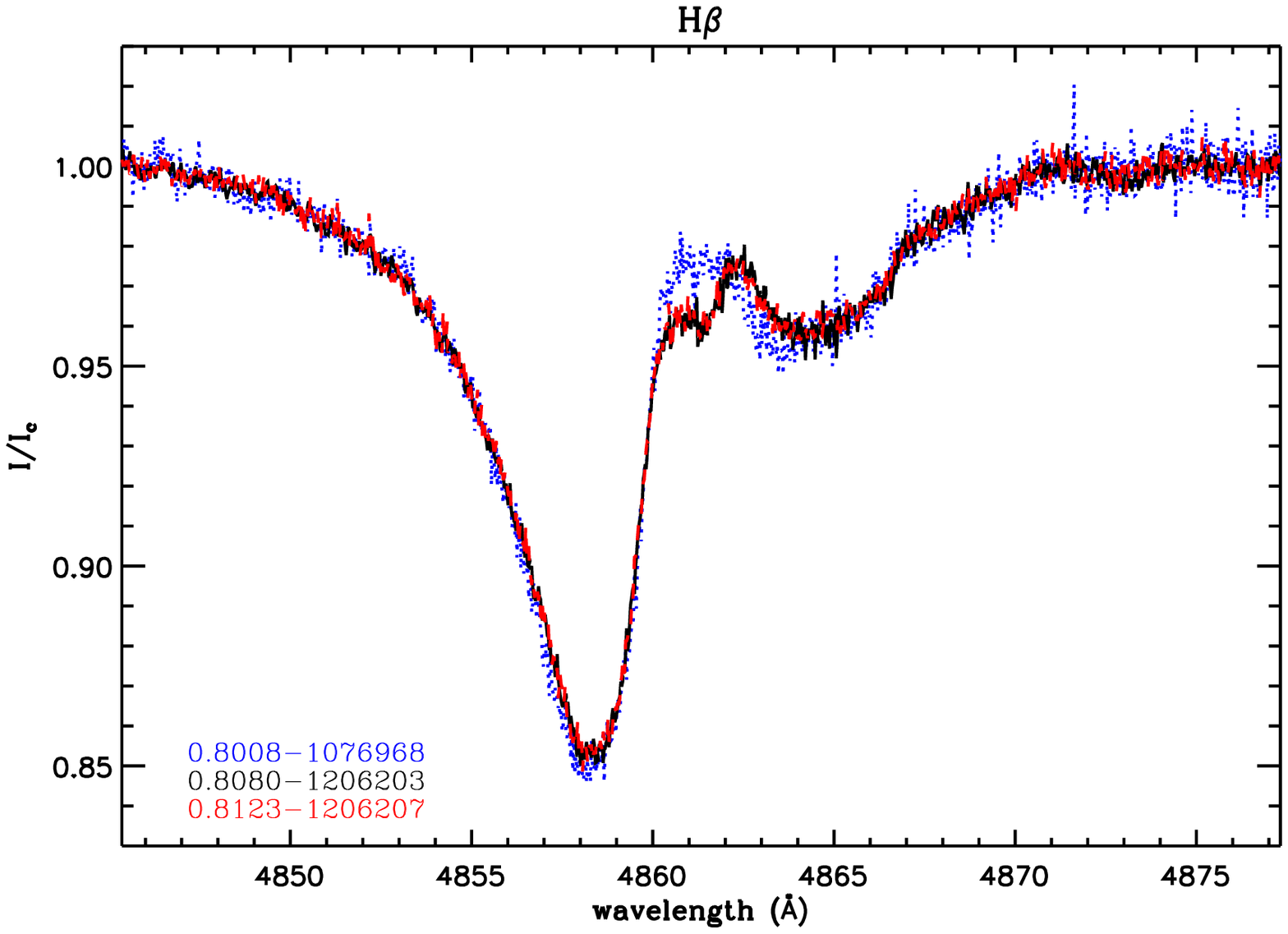}
\includegraphics[width=5.5cm]{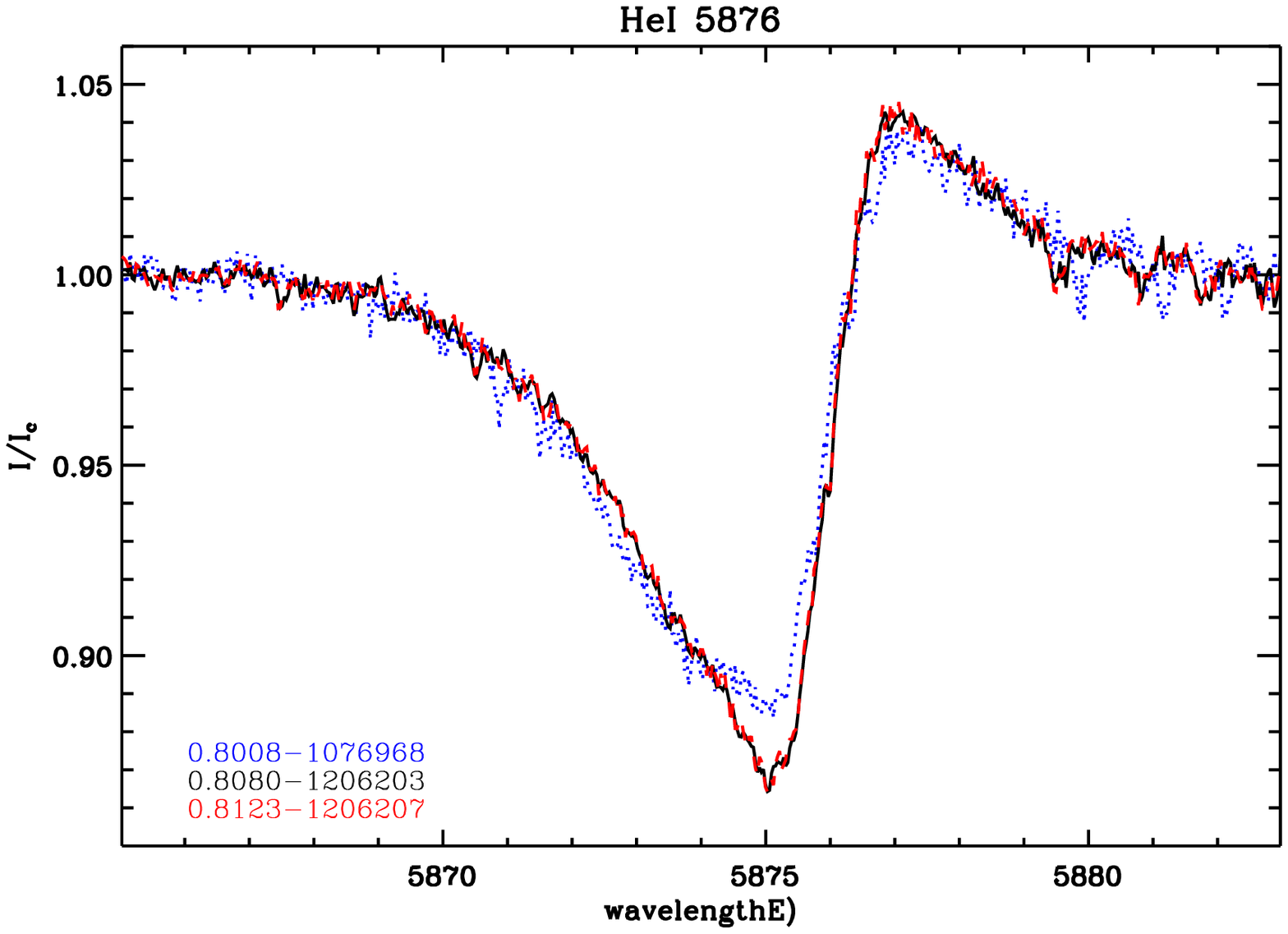}
\includegraphics[width=5.5cm]{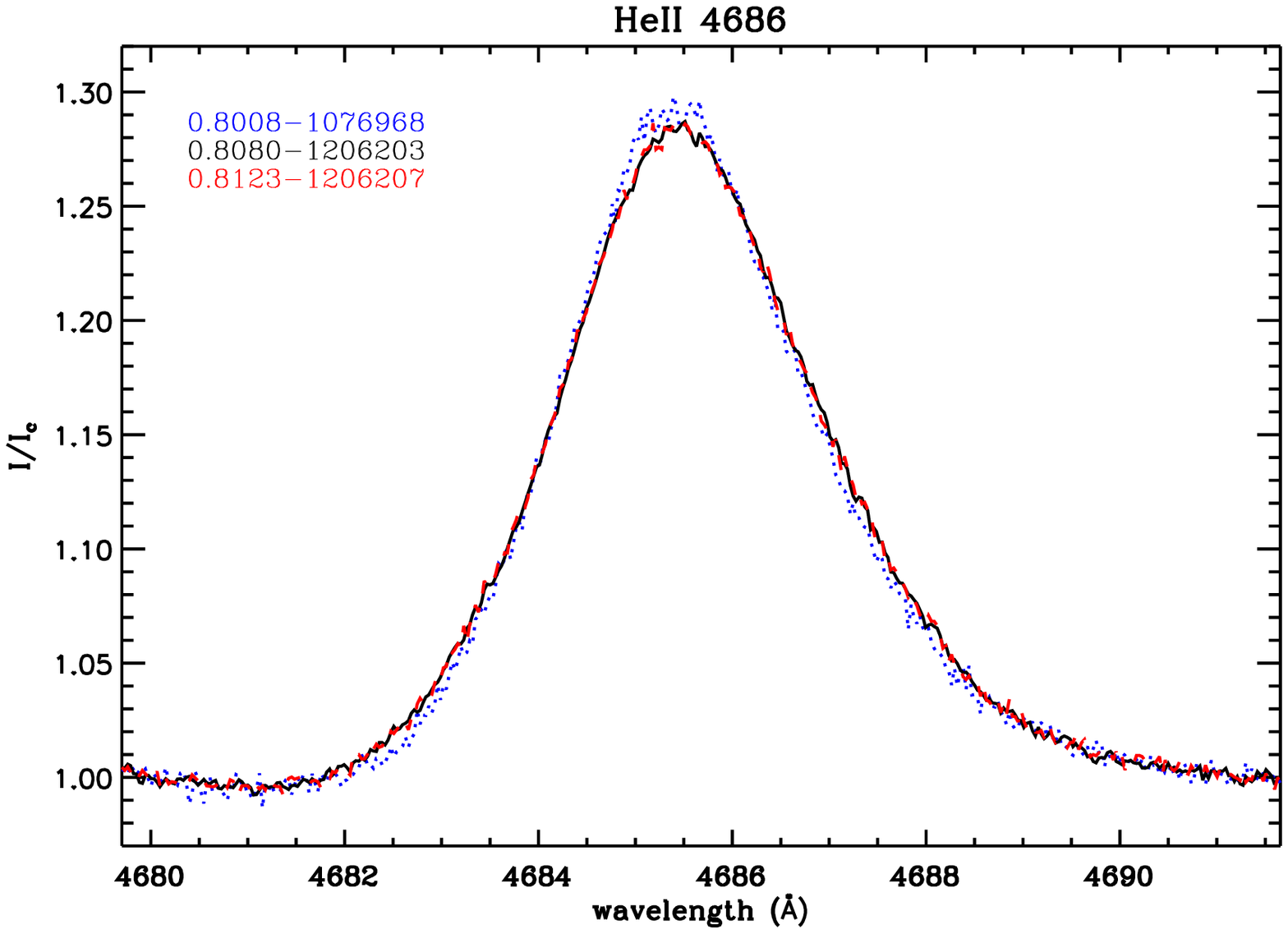}
\includegraphics[width=5.5cm]{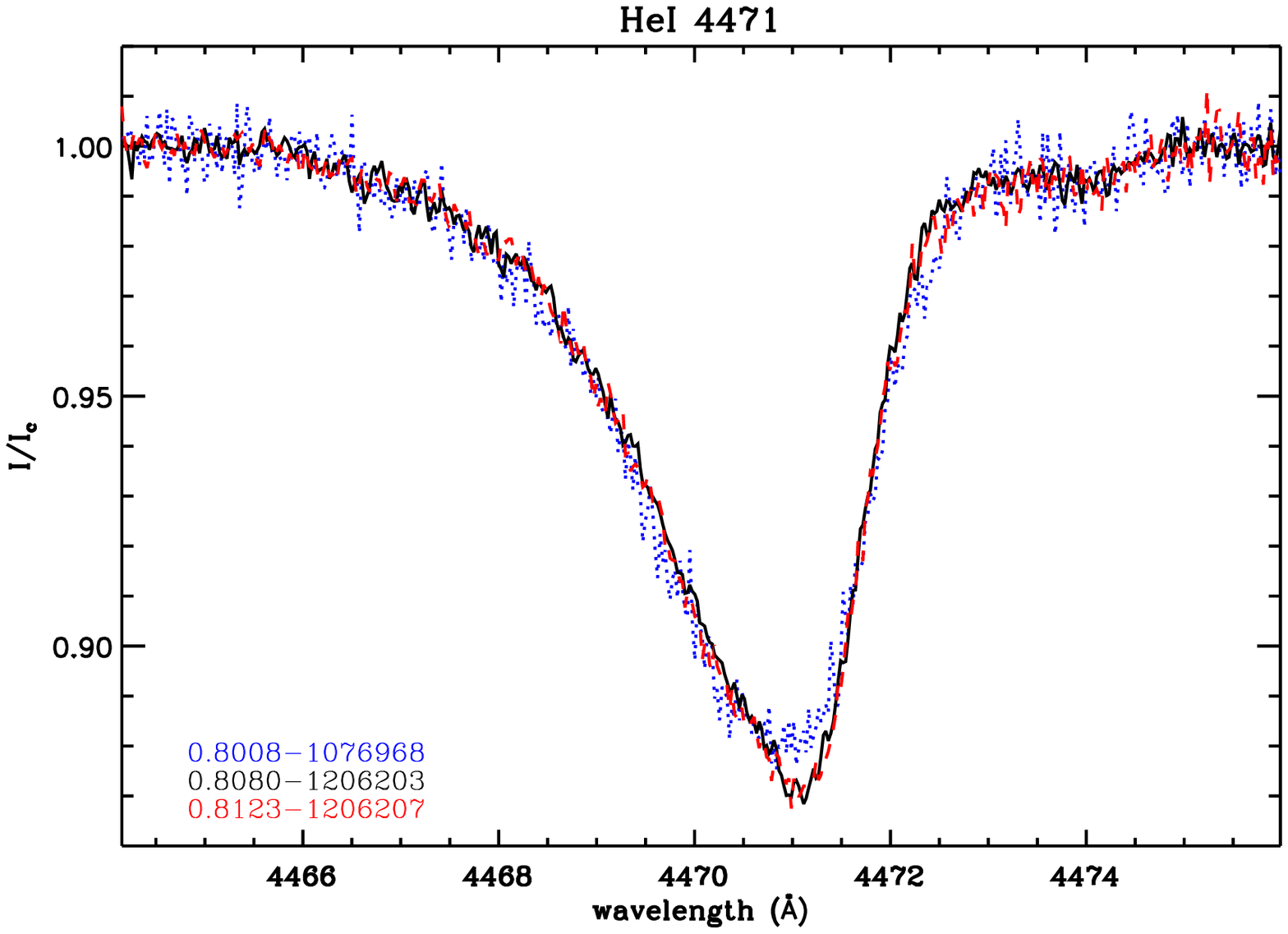}
\includegraphics[width=5.5cm]{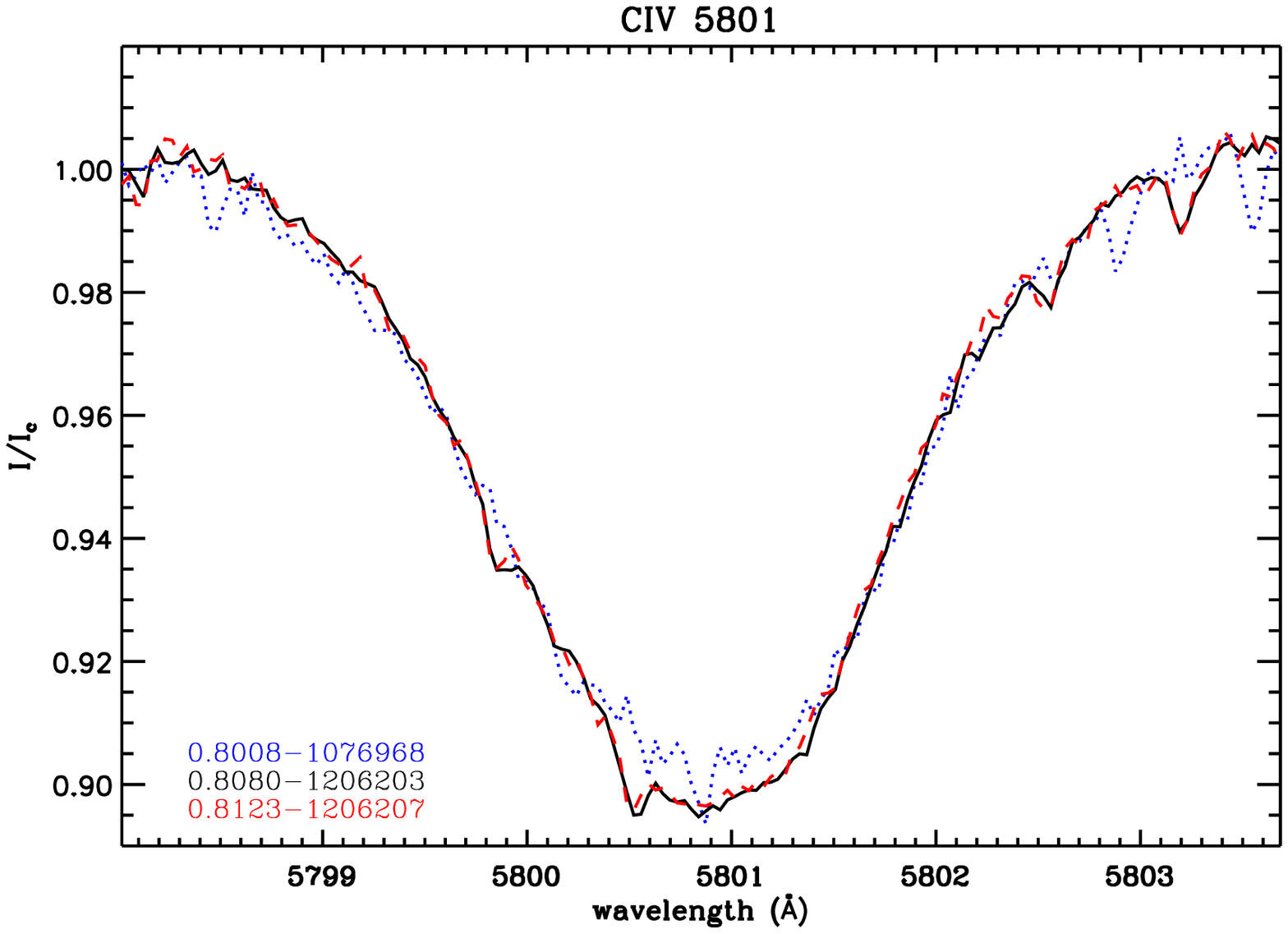}
\caption{Cycle-to-cycle changes of the of H$\alpha$, H$\beta$, He~{\sc i} $\lambda$5876,  He~{\sc ii} $\lambda$4686, He~{\sc i} $\lambda$4471, C~{\sc iv} $\lambda 5801$ lines of HD 148937.  Shown are profiles obtained at nearly identical phases according to Eq. (2). Profiles acquired close in time to each other (\#1206203 - phase 0.8080 - and \#1206207 - phase 0.8123) match perfectly, while a profile obtained 408 days earlier (approximately 58 cycles earlier) at almost identical phase (spectrum \#1076968 - phase 0.8008) shows systematic differences unattributable to local normalisation.}
\label{Profs}
\end{figure*}

\section{Magnetic field}

\subsection{Diagnosis from nightly CFHT spectra}

Least-Squares Deconvolution (LSD, Donati et al. 1997) was applied to all CFHT observations. In their detection of the magnetic field of HD 191612 in 2006, Donati et al. developed and applied an LSD line mask containing 12 lines. Given the similarity between the spectra of HD 148937 and HD 191612, we began by using this line list to extract, for all collected spectra, mean circular polarization (LSD Stokes $V$), mean polarization check (LSD $N$) and mean unpolarized (LSD Stokes $I$) profiles. All LSD profiles were produced on a spectral grid with a velocity bin of 36 km\,s$^{-1}$, providing reasonable sampling of the observed mean profile and enhancing the S/N per LSD pixel. The LSD Stokes $I$ profile shows a strong extension to high ($\sim 350$~km\,s$^{-1}$) velocities in the blue wing. This asymmetry is similar to that observed in the He~{\sc i} lines of HD~148937, and likely is reflective of the inclusion of such lines in the mask. An additional dip and contribution to the depression in the blue wing is attributed to the DIB located at $\sim 579.8$~nm, which was not included in the line mask but which is blended with C~{\sc iv} $\lambda$5801 (which is itself included).

Using the $\chi^2$ signal detection criteria described by Donati et al. (1997), we evaluated the significance of the signal in both the Stokes $V$ and $N$ LSD profiles. We obtain a marginal detection (i.e. a false alarm probability (fap) below $10^{-3}$) in one spectrum (\#1206203). In no case is any signal detected in the $N$ profiles extracted from the individual spectra. If we coadd the LSD profiles of spectra acquired during a single night (typically 2 spectra per night), we obtain one definite detection of signal (false alarm probability ${\rm fap}<10^{-5}$) in the $V$ profiles (for the coadded profile corresponding to spectra IDs 1217692 and 1217696),  and one marginal detection (1218069 and 1218073). An examination of this LSD profile shows a weak asymmetry (positive in the red wing, negative in the blue wing) that is not present in $N$.

To quantitatively characterise the magnetic field implied by our data, from each set of LSD profiles we measured the mean longitudinal magnetic field in both $V$ and $N$ using the first moment method (Rees \& Semel 1979) as expressed by Eq. (1) of Wade et al. (2000), integrating from -280 to +180 km\,s$^{-1}$ (as did Wade et al. 2011 for HD 191612). The longitudinal field measured from Stokes $V$ is detected significantly (i.e. $|z|=|B_\ell|/\sigma\ge 3$) in 3 of our observations. In no case is the longitudinal field significantly detected in $N$. The results of this analysis are summarised in Table 1.

Due to the relatively large width of the profile, and the limited S/N of our observations, our longitudinal field error bars are relatively large: from about 100 G, with a median of 138 G. As expected, the longitudinal field measured from the $N$ profiles is consistent with the absence of any signal in the $N$ spectrum. The inferred field values are distributed randomly around zero in a manner that is statistically consistent with a Gaussian distribution (as characterised by the detection significance $z$). The longitudinal field measured from the $N$ profiles exhibits a mean value of $62\pm 26$~G ($2.4\sigma$). On the other hand, the longitudinal field measured from the $V$ profiles is generally negative, with a (weighted) mean value of $-143\pm 26$~G (5.5$\sigma$). This again suggests the presence of a signal in $V$ that is not detected in $N$.


As a further test, we have investigated if the distribution of longitudinal field detection significance from Stokes $V$ is in fact significantly different from that derived from $N$. We employed a two-sided Kolmogorov-Smirnov test. For the cumulative distributions of $z_V$ and $z_N$, we obtain $D=0.5625$, indicating that the null hypothesis (that the distributions are the same) can be rejected at over 99.9\% confidence.

We therefore conclude that the data provide strong evidence of the presence of a magnetic field in the photospheric layers of HD 148937.

\subsection{Diagnosis from coadded CFHT spectra}

Because of the magnitude of the mean longitudinal field ($\sim 150$~G) and that it appears to remain consistently negative, it is reasonable to assume we are seeing an organised magnetic field, viewed predominantly from a single magnetic hemisphere (i.e. near the negative (south) magnetic pole). In this case, even if the star is rotating according to the $\sim 7.03$~d period, the Stokes $V$ Zeeman signature's shape would vary relatively little from one observation to the next. We therefore proceed to average all LSD profiles to obtain a single grand average profile, with the hope that we are able to detect a significant Stokes $V$ signature.

The grand average LSD profiles are shown in Fig. 7. The $N$ and $V$ profiles are characterised by an LSD noise level (per 36 km\,s$^{-1}$ pixel) of $2.4\times 10^{-5}$. A clear variation is visible in Stokes $V$, whereas $N$ shows no variation whatsoever. The detection probability of a signal within the line in Stokes $V$ is effectively 100\% (i.e. fap of $1.1\times 10^{-8}$), corresponding to a definite detection. No detection is obtained in $N$ (fap of $1.2\times 10^{-1}$). The longitudinal field inferred from the grand average $V$ profile is $-193\pm 26$~G (7.4$\sigma$), while no similar field is detected in $N$ ($37\pm 26$~G; 1.4$\sigma$). The detection of a clear signature in the grand average profile provides clear confirmation that HD 148937 hosts a magnetic field, with a strong disc-averaged and phase-averaged longitudinal component.

We have also searched the grand average spectrum for Zeeman signatures in individual line profiles. As illustrated in Fig. 8, marginal signatures are visible in H$\beta$, He~{\sc i} $\lambda 5411$ and C~{\sc iv} $\lambda 5801$, while a rather strong signature is visible in He~{\sc i} $\lambda 5876$. The shapes of these signatures are compatible with that of the Stokes $V$ LSD profile. Investigation of individual and phase-binned (see below) spectra provided no detection of signatures in individual lines - they are only detectable in the grand average spectrum.

To test the sensitivity of these results to the detailed mask composition, we follow the procedure employed by Wade et al. (2011) for HD 191612, extracting LSD profiles for two additional line masks. We began using a generic line mask based on a 40000~K {\sc extract stellar} request from the VALD database, using a line depth threshold of 0.1. We sought to construct a mask that yielded a mean LSD Stokes $I$ line profile that exhibited as little variability as possible, and that was as symmetric as possible, while still maximizing Stokes $V$ signal. At each step of the mask development, we visualised the agreement of the LSD model (i.e. the convolutions of the Stokes $I$ and $V$ LSD profiles with the line mask) with the reduced spectrum, and evaluated the symmetry and variability of the LSD profile. After removal of the most significantly affected lines, we were left with a mask containing 26 lines, for which we adjusted the mask depths to best match the mean observed line depths. This mask is dominated by lines of neutral He, but also contains lines of ionised He, C, N O and Si. LSD profiles extracted using this mask - which contains more than twice the number of lines in the Donati et al. mask for HD 191612 - yielded LSD profiles nearly identical to those obtained from the Donati et al. mask. This is likely a consequence of the inclusion of many of the same strong He lines in both masks. Measurements of the longitudinal fields extracted from these LSD profiles are in good agreement with those obtained above. 

\begin{figure}
\centering
\includegraphics[width=8cm,angle=0]{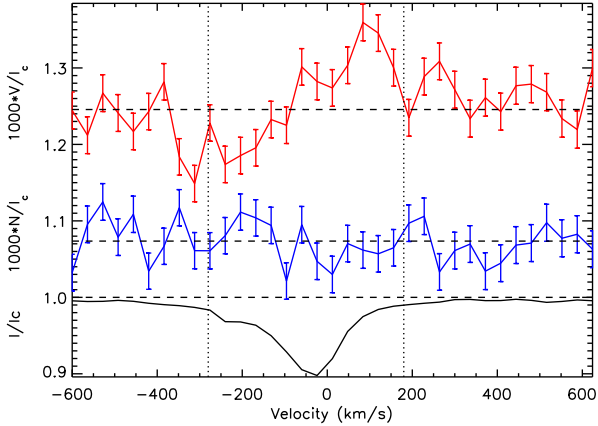}
\caption{Grand average LSD profiles of HD 148937. From these profiles we obtained a definite detection in $V$, and no detection in $N$. The longitudinal field measured from the Stokes $V$ profile within the indicated integration range is $B_{\rm z}=-193\pm 26$~G ($7.4\sigma$). For comparison, the longitudinal field inferred from the diagnostic $N$ profile is $37\pm 26$~G ($1.4\sigma$). The false alarm probability in Stokes $V$ is $1.1\times 10^{-8}$, while in $N$ it is $1.2\times 10^{-1}$.}
\label{SNRs}
\end{figure}

\begin{figure}
\centering
\includegraphics[width=7.5cm,angle=-90]{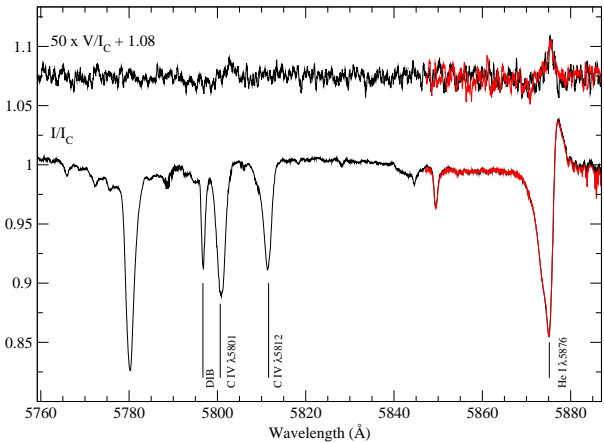}
\caption{Profiles of individual lines in the grand average spectrum of HD 148937. Note in particular the clear Stokes $V$ signature in the absorption component of He~{\sc i} $\lambda 5876$ reproduced in two adjacent orders, and a compatible but much weaker signature in the C~{\sc iv} $\lambda 5801$ line.}
\label{SNRs}
\end{figure}

\subsection{Further tests}

As a second test, we extracted LSD profiles using a more restricted mask containing only a half-dozen metallic lines (lines of  C, O and Si). This mask yields profiles whose shapes differ substantially from those obtained from the first two masks, as is expected due to the exclusion of the much broader He lines. Longitudinal fields computed using this mask (measured using an integration range of -175 to +115 km\,s$^{-1}$) are in formal agreement with those obtained using the masks containing He lines, with a mean of $-160\pm 23$~G. The grand average metallic line LSD profile yields a marginal detection in Stokes $V$ (and no detection in $N$), and corresponds to a longitudinal field of $-121\pm 23$~G ($5.3\sigma$).




\begin{figure}
\centering
\includegraphics[width=8cm,angle=0]{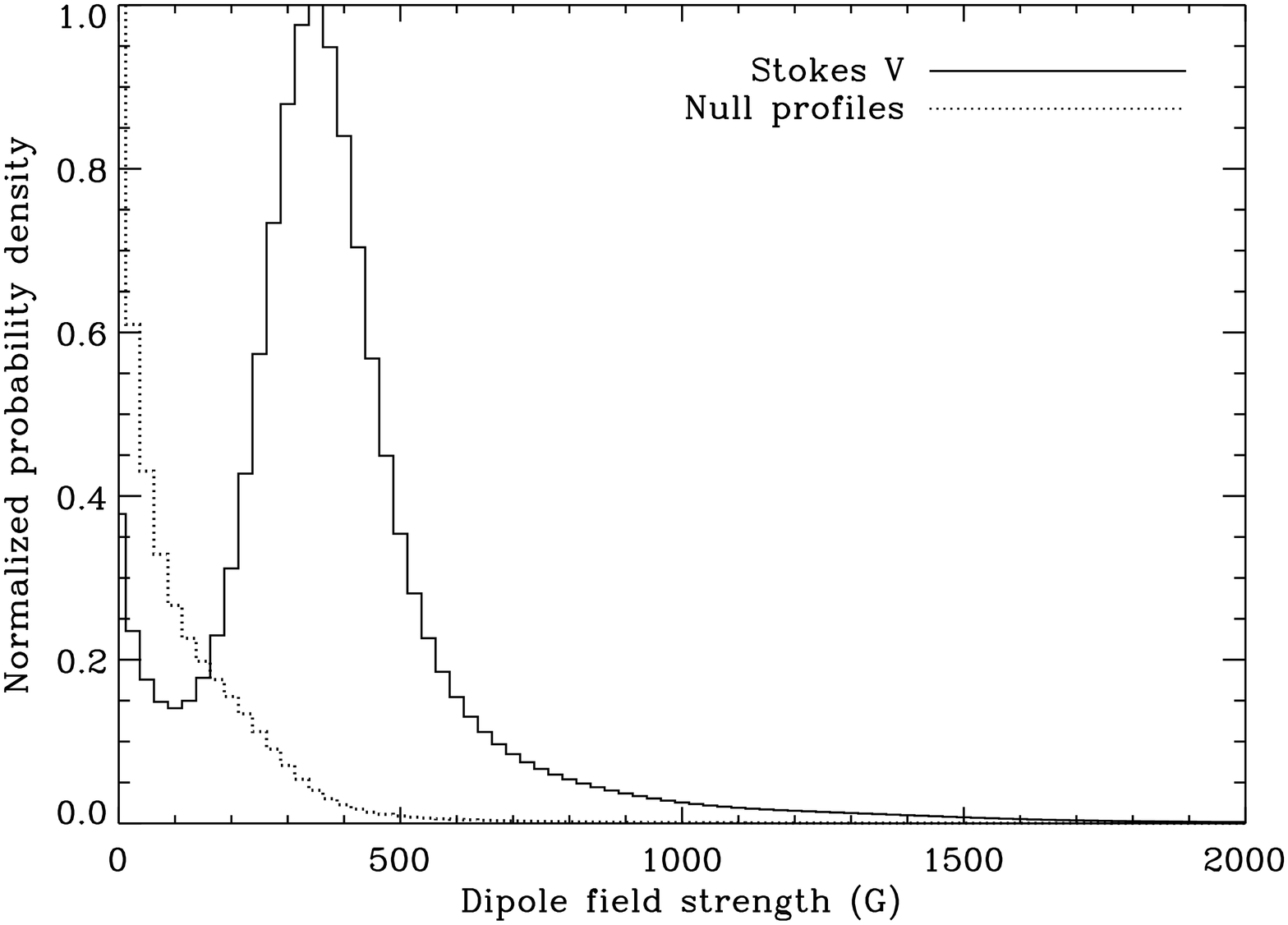}
\caption{The probability density functions marginalised for $B_{\rm d}$ have been normalized by their peak values in order to facilitate graphical representation. The parameter evaluation for the dipole model treat any difference with the model as additional Gaussian noise that is marginalized, leading to the most conservative estimate of the parameters.}
\label{SNRs}
\end{figure}

%

As a third test, we compared the nightly-averaged profiles produced by the metallic line mask to a grid of synthetic Stokes $V$ profiles using the method of Petit \& Wade (2011). We restrict ourself to the metallic line mask for this test because in order to treat the LSD profiles as a single line profile for modelling, it is best to use lines with similar shapes. The synthetic Stokes V profiles were produced by a simple algorithm performing a disk integral and assuming the weak field approximation to compute the local Stokes V profiles. The parameters of the synthetic line were chosen to fit the intensity LSD profile. The magnetic model is a simple centered dipolar field, parametrized by the dipole field strength $B_{\rm d}$, the rotation axis inclination $i$ with respect to the line of sight, the positive magnetic axis obliquity $\beta$ and the rotational phase $\varphi$. Assuming that only $\varphi$ may change between different observations of the star, the goodness-of-fit of a given rotation-independent ($B_{\rm d}$, $i$, $\beta$) magnetic configuration can be computed to determine configurations that provide good posterior probabilities for all the observed Stokes V profiles, in a Bayesian statistic framework. Notice that in order to stay general, we do not constrain the phases nor the inclination. The Bayesian prior for the inclination is described by a random orientation ($p(i)\propto \sin(i)\,di$), the prior for the dipolar field strength has a modified Jeffreys shape with a cut at $B_{\rm d}=40$\,G corresponding to twice the grid step in order to avoid a singularity at $B_{\rm d}=0$\,G. The obliquity and the phases have simple flat priors. In order to assess the presence of a dipole-like signal in our observations, we compute the odds ratio of the dipole model with the null model (no magnetic field, Stokes V = 0). We also perform the same analysis on the null profiles. As expected, no single observation is strongly in favour of the magnetic model, although most of the observations that do favour the magnetic model are clustered around the same phases. Taking into account all the observations simultaneously, the odds ratio is 22:1 in favour of the magnetic model. For $N$, the combined odds ratio is 2:1 in favour of the null model. Note that as the case $B_{\rm d}=0$\,G is included in the magnetic model, the difference between the two models for noise is expected to be dominated by the ratio of priors, i.e the Occam factor that penalizes the magnetic model for its extra complexity. The derived probability distribution functions (PDF) marginalised for the dipole strength are shown in Fig. 9. The most probable dipole strength according to the metallic line Stokes $V$ profiles is about 350~G - approximately 3 times greater than the longitudinal field inferred from those profiles.


Based on this analysis, we conclude that the shapes of the LSD profiles depend on the composition of the line mask. However, all analyses, independent of the line mask used, imply the existence of a magnetic field with a magnitude of the longitudinal component of 100-200~G. Given these results, and for consistency with the analysis of HD 191612 by Wade et al. (2011), we have based our following analysis on the LSD profiles extracted using the 12 line mask of Donati et al. (2006). 

\subsection{Longitudinal magnetic field variation}

To study the magnetic field geometry, we model the longitudinal field variation of HD 148937 as a function of rotational phase. To this end, we proceed as follows. First, we adopt the oblique rotator model as the framework in which we interpret the magnetic observations. This implies that the 7.03 d period (and in particular the ephemeris as expressed by Eq. 2) corresponds to the stellar rotational ephemeris. We have binned the LSD profiles in 7 phase bins (phases determined according to Eq. 2) typically spanning 0.03-0.04 cycles (but as large as 0.06 cycles), and re-measured the longitudinal field from these averaged LSD profiles. The binned $V$ longitudinal field measurements are consistently negative, and a majority of them correspond to detections at 3$\sigma$ confidence or greater. In contrast, the associated measurements obtained from the $N$ profiles scatter randomly about zero. (While one of the $N$ longitudinal field measurements is formally significant at over 3$\sigma$, no corresponding coherent signal is apparent in the $N$ profile). The resultant uncertainties are reasonably uniform, about 60-75~G. These binned results are reported in Table 4. 

\begin{table}
\caption{Longitudinal magnetic field of HD 148937, binned in phase according to Eq. (2). $P_{\rm V}$ and $P_{\rm N}$ are the detection probabilities derived from the LSD Stokes $V$ and $N$ profiles, respectively.  $z=B_{\rm z}/\sigma$ is the detection significance of the longitudinal magnetic field.}
\begin{center}
\begin{tabular}{crrrrrr}\hline\hline
Phase  &   \multicolumn{3}{c}{Stokes $V$} & \multicolumn{3}{c}{Null $N$}\\
             & $\langle B_{\rm z}\rangle \pm \sigma_{\rm B}$ & $z_V$ & $P_V(\%)$ & $\langle B_{\rm z}\rangle \pm \sigma_{\rm B}$ & $z_N$& $P_N(\%)$\\
           & &                             &          &               (G)                                                                \\
\hline
0.081 & $   -189   \pm  63 $ &  -3.0 & 99.906  &   $  -52      \pm  63    $  & -0.8  & 04.703\\   
0.214 & $    -96   \pm  74 $ &  -1.3  & 00.129  &   $  156      \pm  74    $  &  2.1  & 82.031\\   
0.355 & $   -119   \pm  61 $ &  -2.0  & 46.803  &   $  -74      \pm  60    $  & -1.2  & 96.595\\  
0.501 & $   -218   \pm  65 $ &  -3.3  & 70.384  &   $   80      \pm  65    $  &  1.2 & 15.743\\ 
0.662 & $   -285   \pm  74 $ &  -3.8  & 99.087  &   $  254      \pm  74    $  &  3.4  &96.646\\   
0.793 & $   -296   \pm  77 $ &  -3.8  & 38.587  &   $    8      \pm  77    $  &  0.1  & 66.664\\   
0.931 & $   -183   \pm  75 $ &  -2.4  & 99.999  &   $  -50      \pm  74    $  & -0.7  & 87.894\\ \hline\hline
\end{tabular}
\end{center}
\end{table}


%

%

%




The phase variation $B_\ell(\phi)$ of the longitudinal field is illustrated in Fig. 10. A fit by Least-Squares of a cosine curve of the form $B_\ell(\phi)=B_{\rm 0}+B_{\rm 1}\cos(2\pi(\phi-\phi_{\rm 0}))$ to the data yields a reduced $\chi^2$ of 0.3, with parameters $B_{\rm 0}=-200\pm 9$~G, $B_{\rm 1}=90\pm 12$~G and $\phi_{\rm 0}=0.24\pm 0.15$. Therefore, according to Least-Squares, the longitudinal field is significantly different from zero (at over $20\sigma$), and the variation of the field is detected at over $7\sigma$. On the other hand, the phases of the extrema are relatively weakly constrained.  The reduced $\chi^2$ of the data relative to the straight line $B_\ell(\phi)=0$ (the hypothesis of a null field) is 10.2, indicating that the null hypothesis does not provide an acceptable representation of the data. While the reduced $\chi^2$ of the cosine fit to the data is relatively low, it is not unreasonable given the small number of degrees of freedom. 


For illustration, we include the 4 measurements reported by Hubrig et al. (2008) and Hubrig et al. (2011)  in Fig. 10, for both their full spectrum and hydrogen line results. Those measurements are qualitatively and statistically consistent with the variation we derive. 

\subsection{Surface magnetic field geometry}

With the inferred rotational period and derived radius (from Table 1), HD 148937 should have an equatorial rotational velocity $v_{\rm e}=108$~km\,s$^{-1}$. Based on the upper limit on $v\sin i$ and its uncertainty obtained in Sect. 3 (see Table 1), we obtain $i\leq 30\degr$. We therefore conclude that HD 148937 is viewed relatively close to a rotational pole, a conclusion consistent with the proposal of Naz\'e et al. (2010). 

Adopting $i=30\degr$ (which ultimately will yield a lower limit on the inferred magnetic field strength), we have fit the phase-binned longitudinal field measurements with synthetic longitudinal field variations corresponding to a grid of oblique dipole surface magnetic fields characterised by the obliquity angle $\beta$ and the polar field strength $B_{\rm d}$. The best-fit model (according to the $\chi^2$ statistic) is characterised by $B_{\rm d}=-1020$~G and $\beta=38\degr$. The 1$\sigma$ uncertainty contours (shown in Fig. 11) permit models with $B_{\rm d}$ in the range $-700$~G to $-1400$~G and $\beta$ from 20 to $65\degr$. At $3\sigma$ the lower limit on the inferred dipole strength is still 450~G, i.e. the presence of a magnetic field remains confident. We have also performed the same procedure using the unbinned longitudinal field measurements, and we obtain results that are statistically identical. 

The sum of the inclination and obliquity angles for models able to reproduce the observed variation approximately obeys the general rule $i+\beta\simeq 70\degr$. 

The extrema of the best-fit dipole model occur at phases 0.24 and 0.74, with approximate 1$\sigma$ uncertainties of $\pm 0.15$ cycles. On the other hand, the emission line variations (in particular those of H$\alpha$ show extrema at phases 0.0 and 0.5. This will be discussed further in Sect. 7.

\begin{figure}
\centering
\includegraphics[width=7cm,angle=-90]{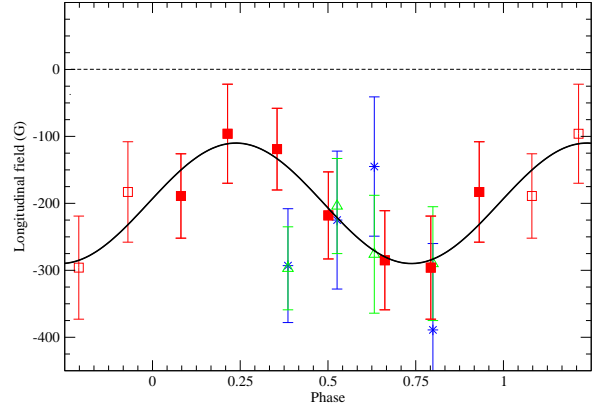}
\caption{Longitudinal magnetic field variation according to the 7.032 d period. Square (red) symbols correspond to the ESPaDOnS observations, while other symbols are the measurements reported by Hubrig et al. (2008, 2011; {\em green triangles = hydrogen lines, blue stars = full spectrum}).}
\label{SNRs}
\end{figure}

\begin{figure}
\centering
\includegraphics[width=8cm,angle=0]{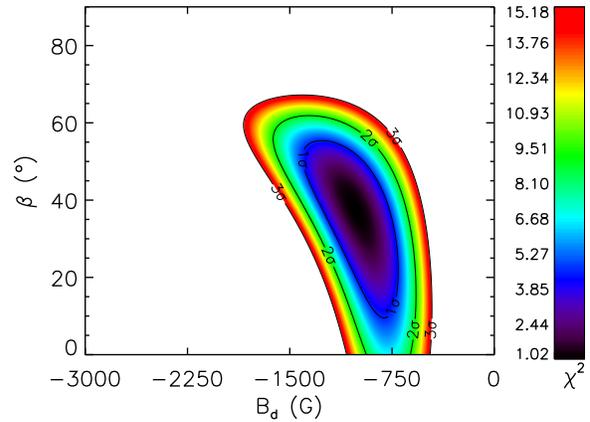}
\caption{$\chi^2$ landscape of dipole field strength $B_{\rm d}$ versus obliquity $\beta$ permitted by the longitudinal field variation of HD~148937, assuming $i=30\degr$.}
\label{SNRs}
\end{figure}

\section{Stellar geometry according to the H$\alpha$ modulation and Hipparcos lightcurve}

We have also attempted to constrain the stellar geometry using the equivalent width variation of the H$\alpha$ line.
We have used the "toy" model of Howarth et al. (2007) which assumes an oblique, rigidly rotating, geometrically thin, optically thick H$\alpha$ disc
that introduces modulation of the H$\alpha$ emission due to its varying projected area as the star rotates. The model includes limb darkening and
inner/outer disc radii that are specified input parameters.

We have computed a number of grids of models, varying the inner and
outer disk radii (inner:outer$=$1.05:1.10, 2.0:2.5, 1.05:2.5 in units of $R_*$) and
linear limb darkening (from 0-1), for rotational axis inclination $i$ from 0-90$\degr$ and disc obliquity angle $\alpha$ (which we
assume to be equivalent to the magnetic obliquity $\beta$ in the magnetic model) from $0-90\degr$, with
steps of 1$\degr$.

For each model we evaluated the $\chi^2$ relative to the observed variation, identifying the 100 best models, which define the
locus of best-fit $\alpha$ as a function of $i$.  Figure 12 illustrates the distribution of these best-fit models in the $i/\alpha$ plane.

We find that the results are very insensitive
to disc size, and that the modest sensitivity to limb darkening is
readily compensated by selecting different pairs of ($\alpha/i$) values.
The models are also highly degenerate with respect to the angles $i$ and $\alpha$.
This is in strong contrast to the results obtained for HD 191612 (Howarth et al. 2007, Wade et al. 2011).
Why do these models constrain the geometry so poorly compared to 191612?
Because the flat-bottomed section of 191612's H$\alpha$ EW variation gives a
strong constraint (the disk must be seen nearly edge-on for a
significant range of phase) - a factor which isn't available for 148937.

Ultimately, we conclude that for $i\leq 30\degr$ (as implied by the upper limit on $v\sin i$ and the 7.03~d period) the disc obliquity $\alpha\geq 10\degr$, and for $i=30\degr$, $10\degr\leq \alpha\leq 25\degr$. These results are compatible with the constraint derived from the magnetic data. Although this modeling is very approximate, it reinforces the applicability of the ORM and supports the interpretation of the emission line variations as due to the varying projection angle of a plasma disc confined to the magnetic equatorial plane.

A second potential constraint on the geometry is potentially derived from the photometric variability. Naz\'e et al. (2008a) reviewed published analyses of photometric observations of HD 148937, and performed their own analysis of Hipparcos and Tycho photometry of this star. They concluded that no significant change or periodicity is present.  Here we compare the observed Hipparcos $H_{\rm p}$ photometric variation, converted to Johnson $V$-band magnitudes according to Harmanec (1998), with the predictions of Townsend's Monte-Carlo radiative transfer code for simulating light scattering in circumstellar envelopes described by Wade et al. (2011). The models were developed for HD 191612, computed for wind confinement parameter $\eta_*$ (defined in the following section) that differs somewhat from that of HD 148937 ($\eta_*\simeq 50$ versus 20), but with lightcurves computed for the constraint derived from the magnetic field variation of HD 148937, $i+\beta\simeq 70\degr$. Unfortunately, the available photometry is not sufficiently precise to allow us to obtain constraints on the geometry. Nevertheless, the predictions demonstrate that high-precision photometry and broadband polarimetry would enable the determination of the individual values of the angles $i$ and $\beta$.

\section{Discussion and conclusions}

We have presented new spectroscopic and magnetic measurements of the Of?p star HD 148937 based on extensive spectropolarimetric monitoring with the ESPaDOnS spectropolarimeter at the Canada-France-Hawaii Telescope, and spectroscopic monitoring using FEROS at the ESO 2.2m telescope. The observations and their analysis were undertaken within the context of the Magnetism in Massive Stars (MiMeS) Project. 

Using the new spectroscopic observations, in tandem with archival UV and IR fluxes, we perform a careful re-determination of the stellar physical properties, and place a slightly improved upper limit on the projected rotational velocity.  

Analysis of the new and previously published optical spectroscopic observations confirms the variability of a variety of spectral lines, and detects variability of others for the first time. Most, and probably all, variable lines vary in phase with H$\alpha$. (The C~{\sc iv}~$\lambda 5801$ line may appear to vary out of phase by $\sim 0.25$ cycles; however, the formal uncertainly of the phases of the extrema is sufficient large that this cannot be robustly established based on our measurements.) The equivalent widths measured from these lines confirm the variation period reported by Naz\'e et al. (2008a, 2010). We find that the variability of the line profiles and equivalent widths is not strictly periodic, and that the absorption and emission line variability exhibits cycle-to-cycle changes. This is tentatively attributed to evolution of the quantity and distribution of emitting material in the magnetosphere of HD 148937. Such a phenomenon has also been reported for the other well-studied Of?p stars (Howarth et al. 2007, Naz\'e et al. 2008b).

Our new magnetic observations of HD 148937 confirm the existence of an organised magnetic field in the photosphere of this Of?p star. The measurements are consistent with a longitudinal magnetic field of constant sign and weak variability. If we interpret the magnetic field and spectral variability within the context of the Oblique Rotator Model (ORM), then the variability period (corresponding to the stellar rotational period) implies an equatorial rotational velocity of 108~km\,s$^{-1}$. To explain the low inferred $v\sin i$, the rotational axis of HD 148937 must be viewed at relatively low inclination, $i\leq 30\degr$. Rotational modulation at low inclination is able to qualitatively and quantitatively reproduce many of the observed properties of HD 148937, in addition to the low $v\sin i$: a longitudinal magnetic field of constant sign and approximately constant strength, the single-wave nature, sinusoidal shape and low amplitude of the H$\alpha$ EW variation, and the lack of observed photometric variability.

\begin{figure}
\centering
\includegraphics[width=7.5cm,angle=0]{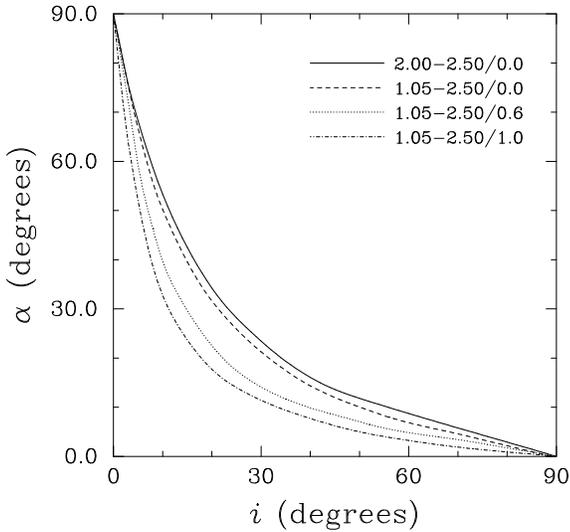}
\caption{Stellar geometry implied by the H$\alpha$ EW variation, according to the model of Howarth et al. (2007). Different curves correspond to differing disc inner-outer radii and $/$limb-darkening coefficients.}
\label{SNRs}
\end{figure}


\begin{figure*}
\centering
\includegraphics[width=16cm,angle=0]{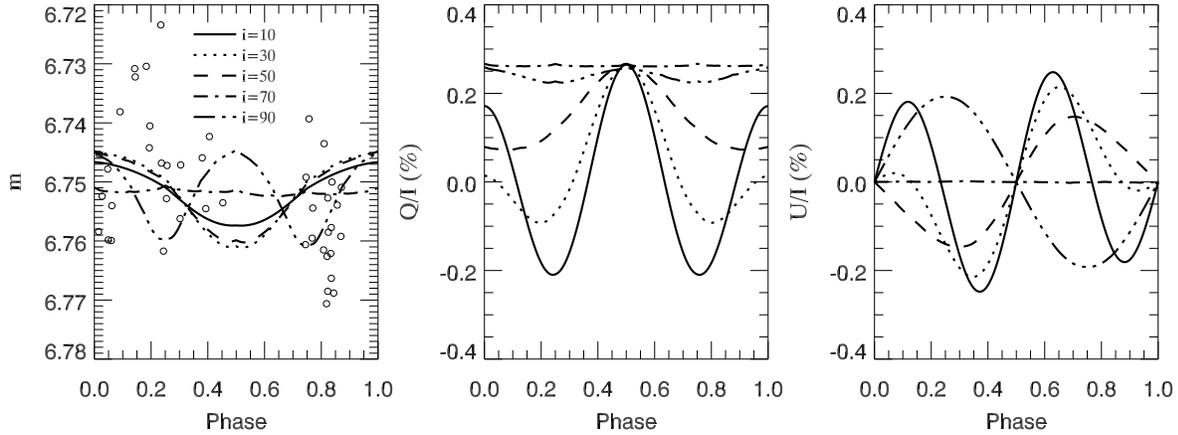}
\caption{Photometric (left) and linear polarisation (centre and right) phase variations of HD 148937 computed for various rotational axis inclinations. The open circles in the left-hand panel represent the Hipparcos photometry converted to Johnson $V$ band. The Hipparcos photometry of HD 148937 is not sufficiently precise to constrain the inclination. Nevertheless, it is clear that high-precision flux and polarisation measurements would yield new constraints on the stellar geometry.}
\label{SNRs}
\end{figure*}

We modelled the longitudinal field variation of HD 148937 according to the dipole ORM, assuming a rotation axis inclination to the line of sight $i=30\degr$ (providing a lower limit on the strength of the (assumed dipolar) surface magnetic field). We derive the dipole polar intensity $B_{\rm d}=1020^{-380}_{+310}$~G and obliquity angle $\beta=38^{+17}_{-28}\degr$. For other values of $i$, the angles follow the approximate relation $i+\beta\simeq 70\degr$. Using the stellar and wind parameters reported in Table 1, we compute the wind magnetic confinement parameter $\eta_*=B_{\rm eq}^2\,R^2/\dot M v_{\rm \infty}\simeq 20$  (neglecting clumping) and rotation parameter $W=v_{\rm eq}/v_{\rm crit}=0.12$. This places the Alfven radius at about $R_{\rm Alf}=\eta_*^{1/4}~R_*=2.0~R_*$. The Kepler (or corotation) radius is located further from the star, at about $R_{\rm Kep}=W^{-2/3}~R_*=4.0~R_*$. As pointed out by ud Doula et al. (2008), for any material trapped on magnetic loops inside the Kepler radius, the outward centrifugal support is less than the inward pull of gravity; since much of this material is compressed into clumps that are too dense to be significantly line-driven, it eventually falls back to the star following complex patterns along the closed field loops. Hence, for HD 148937 (as for HD 191612; Wade et al. 2011), all magnetically-confined plasma (i.e. all wind plasma located inside the Alfven radius) is unstable to this phenomenon. HD 148937 therefore appears to lack a region in its circumstellar environment where centrifugal forces are able to support magnetically-channelled wind plasma against infall back toward the stellar surface (although we recall that the uncertainties on $\eta_*$ are quite large, and therefore the location of $R_{\rm Alf}$ insecure). In any case, while the wind confinement parameter of HD 148937 is comparable to those of the other known magnetic Of?p stars HD 191612 and HD 108 and $\theta^1$~Ori C ($\sim 50$) , its relatively rapid rotation ($P_{\rm rot}\sim 7$~d versus $\sim538$~d (HD 191612), $\sim 55$~y (HD 108) and 15.5 d ($\theta^1$~Ori C) may lead to a greater influence of rotational centrifugal effects on its wind confinement.

The above discussion implicitly assumes that the variable line emission arises in a reasonably static (in the rotating stellar frame) distribution of plasma located principally in the magnetic equatorial plane. We point out that the He\,{\sc ii}\,$\lambda\,4686$ and H$\,\alpha\,\lambda 6563$ line shapes could also be interpreted as P Cyg profiles produced in an outflow. If we assume that these profiles arise from a spherically-symmetric wind (admittedly a rather rough assumption given the observed line variability, and the sensitivity of mass-loss rates determined from recombination lines to the (rather uncertain) volume filling factor), we can use the CMFGEN code to model these lines using the physical parameters in Table 1. We find that the profiles of these lines can be reproduced with an average "density" $\dot M/v_\infty$ of about $1.8\times 10^{-9}$. This is a of factor five larger than that is needed to fit the UV lines ($\dot M/v_\infty=3.8\times 10^{-10}$, neglecting clumping). The width of the emission line profiles however
suggests a substantially lower flow velocity in the line formation region ($\sim 900$~\kms). This results in an only moderately higher mass loss rate ($1.6\pm 0.4\times 10^{-6}~M_\odot$/yr) than inferred by Naz\'e et al. ($1\times 10^{-6}~M_\odot$/yr). As H$\alpha$ is usually formed within one stellar radius above the
stellar surface (i.e. below the Alfven radius in the present case), the line emission may thus originate from a slower, and thus denser outflow in regions of closed magnetic loops.

One peculiarity of the derived magnetic geometry is its phasing relative to the emission line variations. In the case of the Of?p star HD 191612 (Wade et al. 2011), extrema of the H$\alpha$ EW variation occur simultaneously with phases of magnetic extrema. In the case of HD 148937, however, the magnetic extrema appear to be offset from the H$\alpha$ extrema by about 0.25 cycles. In the context of a scattering disc located in the magnetic equatorial plane, this phenomenon is difficult to explain. However, the offset is only a 1.5-2$\sigma$ effect. Given the low amplitude of the longitudinal field variation, we speculate that this is likely an artifact. New, higher-precision measurements are required to clarify this issue.

HD 148937 is distinguished from the other well-studied Of?p stars (HD 108 and HD 191612) by its relatively short rotational period. If we assume that angular momentum is lost through the moment arm generated by the stellar wind and the magnetic field, the characteristic "magnetic braking" spintown time $\tau_{\rm spin}$ (Eq. (25) of ud Doula et al. 2009) is about 1.6 Myr (assuming the characteristics of the UV wind, neglecting clumping), and adopting $k=0.17$ (ud Doula et al. 2009). The spindown time we compute for HD 148937 is somewhat longer than those of HD 191612 and HD 108 (respectively 0.33 Myr and 0.83 Myr using recent physical and magnetic data for those stars). Given the ages of the Of?p stars (2-4 Myr; ud Doula et al. 2009), these differences alone are insufficient to explain the large range of rotational periods of these stars (assuming they all began their lives with similar rotation rates) assuming similar ages. The short rotational period of HD 148937 could therefore imply that this star is significantly younger than HD 108 or HD 191612.

In closing, we note that HD 148937 is surrounded by the bipolar nebula NGC6164-5. It is known to be enriched (e.g. Dufour et al. 1988) and was therefore proposed to be the result of an giant eruption of the star. A larger "bubble" surrounds the system, and is attributed to wind expansion in earlier phases of the star's life (Fairall et al. 1985). The morphology of NGC6164-5 is axisymmetric, with two bright "blobs" to the north and south of the star. While the origin of the nebula is not well understood, such morphologies are often explained by ejection in the same general direction as the stellar rotation axis,. However, the magnetic results presented here rather suggest that this axis is nearly perpendicular to the plane of the sky. A detailed modelling of the nebular ejections should thus be undertaken, taking into account the geometry suggested by magnetic measurements.



The main conclusions of this study are that an organised magnetic field with a dipole surface intensity of approximately 1 kG is detected in HD 148937. Within the context of the ORM, the H$\alpha$ period of 7.03~d is interpreted to be the stellar rotational period. This is, by more than a factor of two, the shortest rotational period found so far for a magnetic O star. The stellar radius, rotation period, $v\sin i$ upper limit, weak longitudinal magnetic field variation, H$\alpha$ equivalent width variation and lack of photometric variability can all be reconciled with the ORM if the star is viewed close to a rotational pole (within $30\degr$). While the detection of a magnetic field in HD 148937 is now relatively robust, the parameters of the oblique rotator model remain rather uncertain - this is a challenging object to observe, and its longitudinal magnetic field intensity and variability are both weak.

Improved geometrical constraints should be obtained by modeling the H$\alpha$ equivalent width variation using more sophisticated models, and measuring and modeling the continuum flux and linear polarisation variation with high precision. In fact, first measurements of the H$\alpha$ line linear polarisation of this star were obtained by Vink et al. (2009), but with no detection of any stellar contribution to the measured polarised flux. (We note however that there was no detection either for $\theta^1$~Ori C, suggesting that higher precision measurements would be desirable).

This is the 3rd Of?p star in which a magnetic field has been firmly detected, confirming that the underlying cause of their spectral peculiarities and variations is almost certainly the magnetic field. We therefore propose that the spectroscopically-identified Of?p stars are themselves a class of magnetic stars - the first among the O-type stars - whose distinctive spectral peculiarities and variability identify them unambiguously as magnetic objects. The test of this proposal will be to search for an unambiguously detect magnetic fields in the two newest members of this class, CPD -28 2561 and NGC 1624-2 (Walborn et al. 2010).

\section*{Acknowledgments}
GAW acknowledges support from the Natural Science and Engineering Research Council of Canada (NSERC). JHG acknowledges financial support from NSERC in the form of an Alexander Graham Bell Canada Graduate Scholarship. RHDT acknowledges support from NSF grants AST-0904607 and AST-0908688. STScI is operated by AURA, Inc., under NASA contract NAS5-26555. { YN acknowledges support from the Fonds National de la Recherche Scientifique (Belgium), the PRODEX XMM and Integral contracts, and the `Action de Recherche Concert\'ee' (CFWB-Acad\'emie Wallonie Europe).}

\end{document}